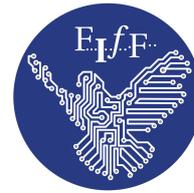

# Data Protection Impact Assessment

# for the Corona App

Kirsten Bock
kirsten.bock@fiff.de

Christian Ricardo Kühne
demian@fiff.de

Rainer Mühlhoff
rainer.muehlhoff@fiff.de

Měto R. Ost
meto.ost@fiff.de

Jörg Pohle
joerg.pohle@fiff.de

Rainer Rehak
rainer.rehak@fiff.de

Version 1.6 – April 29, 2020

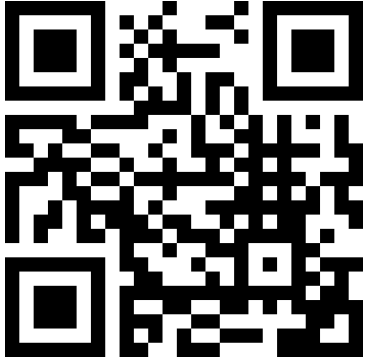

https://www.fiff.de/dsfa-corona

# Contents

# Summary and Findings

Since the SARS-CoV-2 virus started to spread across Europe in early 2020, public and political debates have increasingly centred around a technological solution to this most pressing problem. Could the pandemic possibly be contained by employing tracing apps on everyone's smartphone? These systems would automatically record all users' interpersonal contacts and thus make it possible to quickly trace the infection chains. Then, potentially exposed individuals could be efficiently tracked to isolate them at an early stage of infection.

States such as Singapore, South Korea and Israel have adopted some of the more radical approaches, strategies that represent, from the point of view of European legal systems, disproportionate infringements of fundamental rights. In response, European app initiatives have recently been formed, in particular the Pan-European Privacy Preserving Proximity Tracing (PEPP-PT) consortium, which is seeking to develop a corona tracing app with a commitment to data protection -– or at least to "privacy", which is not the same thing. Thus, there are tracing systems that are currently being designed that are comparatively more data protection friendly than others, even within Europe. For some weeks now, the accompanying media discourse has also conveyed the idea of corona apps *made in Europe* that might respect the "privacy" of all users and comply with the EU General Data Protection Regulation (GDPR).

However, data protection by design is not just a question of implementation, but a more complex consideration in need of precise and detailed discussion. The GDPR itself requires operators of large-scale data processing systems to produce a **data protection impact assessment (DPIA)** if the systems pose high risks to fundamental rights and freedoms. In our report, we show that a corona tracing system falls into this category. A DPIA is a structured risk analysis of data processing – compiled before operation – which identifies and evaluates possible consequences for fundamental rights.

Looking at the planned corona tracing systems, we are dealing with a large-scale social experiment that involves digital behaviour recording under state supervision in Europe. The effectiveness and implications of such apps cannot yet be predicted and it can be assumed that different variants will be rolled out and evaluated across the EU. The consequences in terms of data protection and thus in terms of fundamental rights will potentially not only affect individuals but also society as a whole. For this reason, it would be relevant for the public discussion to both prepare and publish a DPIA. Since none of the parties involved has so far presented a complete and generally accessible DPIA, and even the submitted *privacy impact assessments* are incomplete, we – a group of scientists and data protection experts organised in the NGO Forum InformatikerInnen für Frieden und gesellschaftliche Verantwortung (FIfF, Forum of Computer Professionals for Peace and Societal Responsibility) – wish to present such a data protection impact assessment as a proactive and constructive contribution.

*Summary and Findings*

## Overview of the data processing

In this DPIA, we refer to the primarily discussed frameworks and concept proposals for a European corona tracing app, which are based on near field sensor technology using Bluetooth Low Energy (BTLE). These particularly include PEPP-PT[1], the Decentralized Privacy-Preserving Proximity Tracing project (DP-3T[2]), and a general concept summarised by Chaos Computer Club spokesperson Linus Neumann[3]. Among these projects, the PEPP-PT project offers a framework – i.e. it does not offer a concrete app, but rather a specification for a suitable data processing system. Hence, within this framework, different implementations – i.e. concrete systems/applications – are conceivable; the DP-3T is one of the concrete proposals. The PEPP-PT framework basically allows each European nation to develop its own application. The framework thus strives to offer some national degrees of freedom while ensuring cross-border interoperability.

In this context, a key finding of our investigation is that the frameworks considered here – especially PEPP-PT – **do not specify important technical characteristics and process properties that have serious implications for data protection**. It is possible to roughly differentiate between at least three system architectures, all compatible with the PEPP-PT framework:

a) **A centralised architecture**: The anonymity of users and confidentiality of contact events is only provided with regard to outside entities, i.e. other users or external actors; the operators and involved authorities can, however, identify all users and connect them to recorded contact histories.

b) **A partially decentralised architecture**, which also allows for **epidemiological research** (e.g. DP-3T): Users and contact events are only concealed from other users and third parties; the server can de-anonymise positively tested users. The system also has a data donation function, by which users can choose to share their contact histories for epidemiological research. If done so, positively tested users' contact events would become visible to operators and authorities.

c) **A completely decentralised architecture** (see Neumann 2020): Users remain anonymous towards other users and third parties and their contact events would also remain secret. Operators and authorities can de-anonymise positively tested users but not their contact history. Epidemiological research is not possible.

| Operators and authorities can ... | Type a | Type b | Type c |
| --- | --- | --- | --- |
| **... de-anonymise all users** | yes | no | no |
| **... de-anonymise positively tested users** | yes | yes | yes |
| **... de-anonymise all contact events** | yes | partly | no |

---

[1] Pan-European Privacy-Preserving Proximity Tracing (PEPP-PT) (2020). URL: `https://www.pepp-pt.org/` (visited on 04/08/2020)

[2] Carmela Troncoso et al. (2020). *Decentralized Privacy-Preserving Proximity Tracing*. White Paper Version: 10th April 2020

[3] Linus Neumann (2020). *"Corona-Apps": Sinn und Unsinn von Tracking*. URL: `https://linus-neumann.de/2020/03/corona-apps-sinn-und-unsinn-von-tracking/` (visited on 04/09/2020)

Our **DPIA mainly refers to the most data protection friendly type, type c**, however, where useful, we also take the technical details of type b into consideration.

First, we conclude that **even the decentralised version entails serious vulnerabilities and risks** that have to be dealt with (for details see our full DPIA). Second, a comparison of the centralised and decentralised types shows that **essential data protection properties depend on the decision to opt for either a centralised or decentralised version.** Therefore, a PEPP-PT compliant version would not necessarily be data protection friendly.

**Important insights, risks and solution approaches**

We now summarise some key insights and outline some risks and solution approaches:

1. **The often stressed voluntariness of the app is illusory.** The use of the app as a prerequisite for relaxing the current corona quarantine is possible and is even already being discussed. People would then have to present the app before accessing public or private buildings, rooms or events. It is also conceivable that employers might swiftly adapt those practices to reopen their establishments more quickly by implementing such voluntary protective measures. This scenario would lead to *implicit coercion to use the app* and result in extremely unequal treatment of those who do not use the app. Since not everyone owns a smartphone, this would also discriminate against already disadvantaged groups.

2. **Without a capacity to intervene and a narrow purpose limitation, the protection of fundamental rights will be at stake.** There would be a high risk that mistakenly detected exposure events (false positives, for example, by contact measurements through walls between two apartments) would result in the unjustified imposition of self-isolation or quarantine. To prevent this, there is a need for legal and actual possibilities to effectively intervene, such as the recall of incorrect disease notifications, the deletion of incorrectly registered contact events with an infected person and the possibility to contest restrictions imposed as a result of the data processed. So far, none of the proposed systems account for that.

3. **All proposed tracing systems process personal data concerning health.** The procedure consists of processing contact data on smartphones, the transmission of this data to a server after the diagnosis of an infection and finally the distribution of this data to all other smartphones for testing for possible contacts with infected people. All data on a smartphone is personal data, related to the user of the device. Since only people who test positively transmit data to the server, the data transmitted constitutes data concerning health. The data processing is thus clearly subject to the GDPR.

4. **The anonymity of users must be enforced by a combination of legal, technical and organisational measures.** References to a natural person can only be effectively and irreversibly separated from the processed data through a multidimensional approach so that the resulting data is anonymous. All the proposals lack such an explicit separation process. In this DPIA, we have laid out legal, technical and organisational requirements that can be implemented to ensure effective and irreversible separation in practice – only under these conditions can the resulting non-identifying infection-indicating data (NIID) be stored on the server and distributed to all other apps.

For a comprehensive explanation of the risks and weaknesses please refer to chapter 7 and for the necessary protecton measures to chapter 8 within the DPIA.

Generally, the data protection perspective sees **the essential risks of data processing as relating to the operators of a data processing system.** Therefore, it is strictly necessary that the barriers to improper processing – e.g. processing exceeding the purpose of the data processing – consist of an effective mix of legal, technical and organizational measures – and do not just entail the operators paying lip service to data protection. Measures taken must be actively made verifiable and documented clearly.

Open source development of the server and app software including all components – for example, in the form of free software – is an essential prerequisite for transparency regarding the implementation of data protection principles. This applies to data protection supervisory authorities but especially also to the data subjects and all members of (civil) society. Only then can confidence and trust be established among people not familiar with the details of information technology.

Third parties may also pose risks to fundamental rights. This does not primarily refer to hackers, but more to commercial actors like large platform operators or state bodies. These organisations may benefit from an increased volume of available tracking data since, first, the Bluetooth module must be permanently kept active for the corona app and, second, through access to extensive data stored in private actors' silos.

**A data protection analysis considers the entire processing and its context, not only the app used for it.**

In the public discourse and in the app projects we examined here, data protection is still reduced to the protection of privacy, i.e. to maintaining secrecy and confidentiality with regard to operators and third parties and in relation to aspects of IT security like encryption. This limited view not only ignores the considerable social and political fundamental risks we have identified in this impact assessment, it even obfuscates them.

# Chapter 1

# Introduction

According to Art. 35 of the General Data Protection Regulation (GDPR) the controller of a processing has to carry out a data protection impact assessment (DPIA) if the planned processing is likely to result in a high risk to the rights and freedoms of the data subject.

It is yet unclear which organisation will be in control and responsible for design and operation of the data processing that utilizes the tracing app, if ever realized. Neither is the legal and functional design of the data processing yet determined. *This DPIA is therefore based on a data processing using a tracing app which, in the authors' judgment, functionally completely fulfills the purpose while at the same time it infringes the rights and freedoms of persons to the lowest extent possible.* For this purpose we anticipate the potential impact in terms of the relevance of the identified risks (validity) as well as the effectiveness and resilience (reliability) of the measures taken.

In principle, by carrying out and documenting a DPIA a controller practices accountability (Art. 5 para. 2 GDPR) in compliance with data protection requirements.

## 1.1 Goals and purposes of this text

The goal of the present text is the production of a DPIA according to Article 35 GDPR for the processing of contact data using a Smartphone app. The processing comprises the app using near-field sensor technology such as Bluetooth Low Energy (BTLE) Beacons to record contact events. Then, in case of infection of the app user, data is transferred from the app to one or more servers. This data will then be distributed to all users for the purpose of informing them about possible contacts with SARS-CoV-2 infected people. The requirements of Art. 35 GDPR are fulfilled, when firstly the controller receives a DPIA including recommendations, and secondly, when the recommended measures have been implemented as well as proof of their effectiveness has been presented

For the present DPIA we distinguish the following types of smartphone apps according to the processed data types:

– **Type 1:** Processing of location data (GPS data, mobile phone metadata)

– **Type 2:** Processing of movement data (aggregated GPS, mobile phone metadata)

– **Type 3:** Processing of contact data (near field sensors, e.g. Bluetooth)

  ○ **Type 3 "centralized":** server gets all contact events of the infected (including information on IDs of those at risk), server informs people at risk. (Examples: TraceTogether, centralized PEPP-PT interpretation).

  ○ **Type 3 "decentralized + epidemiological":** Server gets the (mathematically related) TempIDs of infected persons, the smartphone calculates

> risk locally and informs users locally. The system has a data donation function, by which users can provide their contact histories for epidemiological research – in these cases contact events are disclosed to the controller and authorities. (Examples: DP-3T, decentralized PEPP-PT interpretation).
>
> ○ **Type 3 "purely decentralized":** Server gets the (mathematically independent) IDs of infected persons, the smartphone calculates risk locally and informs users locally. Epidemiological research is not supported. (Example: Summary of Linus Neumann, cf. Neumann 2020.)

For analysing the whole processing procedure this DPIA primarily considers the processing activity (PA) of data corresponding to type 3. For reasons of focus, one single purpose is set: informing people in risk of infection. The PA considered therefore refers to type 3 "purely decentralised", enriched by technical details of the DP-3T project, when it supports the purpose as set above. The temporary identifiers (*TempIDs*) generated by the users' apps are transmitted via Bluetooth and become *health tempIDs* in case of a COVID-19 diagnosis. They are then sent to the server by the CV-infected person, where they are anonymized by a secure separation procedure ensured by a combination of legal, organizational and technical means. Thereafter, these *non-identifying infection-indicating data* (NIID) can be downloaded by all app users. This enables other users to calculate possible exposure events based on *contact data*, meaning the received "foreign" TempIDs including duration and signal strength profile.

It is not the goal of this DPIA to specify or implement a type 3 app, but exclusively to identify risks and protective measures according to criteria defined by the GDPR.

It is not our goal to carry out a "Privacy Impact Assessment" focussed only on effects regarding privacy or the private sphere, but we analyse effects on all fundamental rights and freedoms of natural persons.

It is not our goal to create a "Surveillance Impact Assessment" only analysing processing forms which can be described as surveillance, but all processing of personal data is considered.

In case of an implementation of this fundamental rights preserving procedure this text intends to be used as formal, structural and substantive preparatory work.

It is essential for a DPIA under the GDPR to not focus on the prominent technology, in this case the "Corona app". Instead, the processing procedure as a whole is in the center of a DPIA, which consists of several processing activities of personal data, in which processes or series of processes take place – partly supported by technology. Considerations must therefore go beyond the use of the app itself, because the app itself is not the whole processing. Essential defining feature for a procedure or for processing is the designated purpose (Hoffmann 1991).

In data protection risk modelling must be based decisively on the risk perceived from the data subjects' perspective. The risk is that the whole processing is not sufficiently designed and operated in a way that it preserves fundamental rights. It is insufficient to exclusively or especially analyse the risks of non-sufficient IT security or the risk of financial losses (Rost 2018b, Bieker, Bremert, and Hansen 2018).

As methodological foundation for the transformation of normative requirements into functional requirements we use the Standard Data Protection Model (SDM) version V2.0a. It was approved by the 98th Conference of the German Independent Data Protection Authorities of the German Federal Government and the German federal states in November 2019 as well as it has been recommended to all public authorities by the IT Planning Council (DSK SDM2.0a).

## 1.2 Target group of the text

The presented DPIA is aimed at

- political decision-makers on the EU level, federal level as well as in federal states,
- Data Protection Supervisory Authorities on the EU level, federal level as well as in federal states,
- technical and legal experts,
- possible manufacturers and operators and
- all other stakeholders.

## 1.3 Members of the project group

**Ass. jur. Kirsten Bock** studied law in Kiel and Guildford/UK with a focus on legal philosophy and logic analysis. She works in the regulatory field and is a member of expert working groups of the European Data Protection Board (EDPB). She conducts research on fundamental ethical and social issues of data protection, certification and the standard data protection model (SDM). She is a member of FIfF.

**Dipl.-Inf. Christian Ricardo Kühne** studied philosophy, computer science and sociology; currently he does research as a free academic in the areas of commons theory and critical computer science concerning emancipatory information and communication technologies. Besides that he is working with the GNUnet project on an alternative internet stack. He is a member of the FIfF.

**Dr. Rainer Mühlhoff** studied mathematics, computer science and theoretical physics, and holds a PhD in philosophy. His research interests include data protection in the context of anonymous mass data, ethical questions of artificial intelligence and social theory of the digital society. He works at the Excellence Cluster Science of Intelligence at the Technical University of Berlin, is co-founder of the Berlin Ethics Lab for Responsible AI and Responsible Human-Computer Interaction and a member of FIfF.

**Dr. Jörg Pohle** studied computer science, law and political science, received his doctorate in computer science with a dissertation on the history and theory of data protection and implications for ICT design. His research concerns ICT analysis and design, legal informatics, and digitisation and sociological theory. He leads the "Data, Actors, Infrastructure" research programme at the Alexander von Humboldt Institute for Internet and Society in Berlin, Germany, and is a FIfF member.

**Dr. Jörg Pohle** studierte Informatik, Rechts- und Politikwissenschaften und promovierte in Informatik zur Geschichte und Theorie des Datenschutzes und Folgerungen für die Technikgestaltung. Er forscht zu Technikanalyse und -gestaltung, Rechtsinformatik sowie Digitalisierung und Soziologischer Theorie. Er ist Leiter des Forschungsprogramms "Daten, Akteure, Infrastrukturen" am Alexander von Humboldt Institut für Internet und Gesellschaft in Berlin und Mitglied des FIfF.

**Dipl.-Inf. Rainer Rehak** studied computer science and philosophy in Berlin, Hong Kong and Beijing. He is currently doing his doctorate at the Weizenbaum Institute for the Networked Society in systemic IT security. He does research and teaching in the fields of data protection and data security, government hacking and technical ascriptions. He is active with Amnesty International, acts as a subject matter expert, for example for parliaments or the Federal Constitutional Court, and he is vice chairman of the board at FIfF.

For questions, criticism, additions or suggestions please contact the authors via `dsfa-corona@fiff.de`

## 1.4 Documents regarding a methodological and substantive foundation for carrying out a DPIA

For a methodological conduction of this DPIA in accordance with Article 35 GDPR we particularly referred to the following documents:

- Article 29 Data Protection Working Party (2013). *Opinion 03/2013 on purpose limitation.* Working Paper WP 203

- Article 29 Data Protection Working Party (2017). *Guidelines on Data Protection Impact Assessment (DPIA) and determining whether processing is "likely to result in a high risk" for the purposes of Regulation 2016/679, as last revised and adopted on 4 October 2017.* Working Paper WP 248

- Konferenz der unabhängigen Datenschutzaufsichtsbehörden des Bundes und der Länder (2018a) [DSK KP5]. *Datenschutz-Folgenabschätzung nach Art. 35 DS-GVO.* Kurzpapier Nr. 5

- Konferenz der unabhängigen Datenschutzaufsichtsbehörden des Bundes und der Länder (2018b) [DSK KP18]. *Risiko für die Rechte und Freiheiten natürlicher Personen.* Kurzpapier Nr. 18

- Konferenz der unabhängigen Datenschutzaufsichtsbehörden des Bundes und der Länder (2019) [DSK SDM2.0a]. *Das Standard-Datenschutzmodell: Eine Methode zur Datenschutzberatung und -prüfung auf der Basis einheitlicher Gewährleistungsziele (English version is available: "The Standard Data Protection Model: A method for data protection consulting and auditing based on uniform warranty objectives").* Version 2.0a. URL: https://www.datenschutz-mv.de/datenschutz/datenschutzmodell/ (visited on 04/21/2020)

- European Data Protection Board (2019b). *Guidelines 4/2019 on Article 25 Data Protection by Design and by Default.* Adopted on 13 November 2019

- European Union (2016) [GDPR]. *Regulation (EU) 2016/679 of the European Parliament and of the Council of 27 April 2016 on the protection of natural persons with regard to the processing of personal data and on the free movement of such data, and repealing Directive 95/46/EC (General Data Protection Regulation), OJ L 119, 4.5.2016, 1–88*

- Michael Friedewald et al. (2017). *Datenschutz-Folgenabschätzung: Ein Werkzeug für einen besseren Datenschutz.* White Paper. Version 3. Forum Privatheit

## 1.5 Acknowledgements

We are especially grateful to Martin Rost for fruitful suggestions, critical feedback and enlightening discussions. In addition we want to thank Malte Engeler, Kai Gärtner, Julian Hölzel, Niklas Rakowski and Lena Ulbricht for valuable exchanges.

For diligent corrections and continuous help in questions of readability concerning the German version we would like to thank Heidi Rehak.

For their excellent translations and corresponding corrections we would like to thank Roisin Cronin (English), Yeimy Carolina González Vargas (Spanish), Amelie Iacono (French), Barbara Kipping (English), Ricardo Mortez (Spanish), Eva Schlehan (English) and Martina Vortel (English).

Last but not least we would like to thank the makers of the technical infrastructure we use, namely Jitsi, LaTeX, Etherpad-lite, Nextcloud, OnlyOffice, Signal, Threema, Sharelatex, DeepL and Zotero.

*Chapter 1 Introduction*

# Chapter 2

# Contextualisation of the processing

The data processing addressed by the present DPIA is related to the global corona pandemic, which began in Wuhan, People's Republic of China, at the end of 2019 and has since spread throughout the world. After the first appearance of infections in Europe at the end of January 2020 – first in France on January 24, four days later in Germany – it took until mid-March – about 40 days – until comprehensive measures were taken in Germany, both preparatory and containment measures. On 22 March 2020, the Federal Government and the Länder agreed on a "comprehensive restraining order", and the Länder of Bavaria, Berlin, Brandenburg, Saarland, Saxony and Saxony-Anhalt agreed on an even further reaching curfew. As of 7 April 2020, the measures are to be maintained until at least 19 April 2020.

Technology-supported processing activities are currently being discussed in public, which should allow 1) to track the spread of the pandemic in order to be able to make predictions about its further spread, 2) to stop the pandemic and at least to control it, 3) to inform potentially infected persons about their possible infection and 4) to monitor and enforce restraining orders, the curfews and/or individual quarantine obligations to move from general measures such as general curfews to risk and target group specific measures to minimise the impact of infection control measures, particularly on the economy.[1]

Technology-supported processing activities globally

In the meantime, a whole range of technology-supported processing activities has been introduced in a large number of countries, including the People's Republic of China, South Korea, Singapore, Israel and Austria.[2]

From mid-February 2020, the People's Republic of China obliged citizens to install and use an app on their smartphone to monitor their movement in order to enforce quarantine measures and identify contact persons. The app transmits movement data to the server(s) where (possible) contacts with infected persons are identified. The smartphone owners are then informed with color codes about how they should behave – from more or less freedom of movement to forced isolation for several weeks. The app shows the status regarding a corona infection, it must be shown at police and other control points.

South Korea uses telephone and credit card data of infected persons to track their previous movements in order to identify possible contact persons. Persons found to have been in the vicinity of infected persons receive telephone calls with information about their previous movements. In addition, the movement data of infected persons are published, according to the government, anonymously, but many of those affected

---

[1] An extensive contextualisation of a processing activity is not an obligatory component of a DPIA according to Art. 35 GDPR. We are discussing other forms of processing activities with the same, similar or neighbouring purposes in order to demonstrate that there is a solution for a narrow legitimate purpose that infringes less intensively on fundamental rights of citizens.

[2] A continuously updated overview can be found at GDPRhub-Liste.

have already been re-identified, and there have been cases of social stigmatisation (Kim 2020).

Singapore's Government Technology Agency launched on March 20 the distribution of the smartphone app "TraceTogether", which is designed to support contact tracing activities of the Ministry of Health. Using the app, users are identified at a central server, then a centrally assigned pseudonym is sent to the app, which is then exchanged via Bluetooth with nearby devices. This data is also sent to the server. In the event that a person is infected, the contacts are identified and the relevant persons are informed in order to send them to quarantine. This quarantine is enforced by monitoring the infected persons using GPS location data of their smartphones; the infected persons must prove that they are actually in their home by providing photos of their living environment upon request.

In Austria, the Austrian Red Cross launched the "Stopp Corona" app on 25 March 2020. On the basis of unique user IDs, the app allows data to be exchanged between smartphones that are close to each other, which are then stored as contact data. In the event of an infection, users should use the app to contact the Red Cross, which will then notify the contact persons with whom they have been in contact in the past 3 calendar days via the app. This notification triggers the collection of further data, including the mobile phone number of the infected person. The Austrian Red Cross explicitly acts as the data controller according to the GDPR. The app is developed and operated by Accenture, the services are hosted in Microsoft's Azure Cloud, and Google's Firebase Cloud Messaging is used for the notifications.[3]

On March 18, 2020, the Israeli government, bypassing the Israeli Parliament, the Knesset, passed regulation that allows the domestic intelligence service, Shin Bet, and the police to monitor mobile phones of confirmed and suspected COVID-19 patients. All mobile phone data that have been collected secretly for years, arguably for counter-terrorism purposes, are now being used to fight the pandemic (Halbfinger, Kershner, and Bergman 2020). These precautions are designed to enforce the comprehensive quarantine measures, both against those whose infections are confirmed and suspected, as well as all those with whom infected persons have had contact in the 14 days preceding their diagnosis. It is not only the location and the movement data that is analysed, but also financial transaction data (Landau, Kubovich, and Breiner 2020).

Technology-supported processing activities in Germany and Europe

In this section, we will briefly introduce five corona (contact) tracing systems, which are in different stages of development, as it is mainly these developments that have or will have a great influence on the situation in Germany: the European projects Pan-European Privacy-Preserving Proximity Tracing (PEPP-PT) and Decentralized Privacy-Preserving Proximity Tracing (DP-3T), the "Corona Data Donation" app of the Robert Koch Institute (RKI) and a proposal by Linus Neumann for a contact tracing app. On 15 April 2020, the EU Commission in cooperation with the Member States and the European Data Protection Board (EDPB) has published a concept containing approaches for a coordinated way out of the current lockdown, which provides for the use of apps for tracing coronavirus infections. In addition, on 10 April 2020, Apple and Google announced a cooperation to develop a contact tracing platform.

The Pan-European Privacy-Preserving Proximity Tracing (PEPP-PT) project, which is yet to be established as a non-profit organisation in Switzerland, intends to propose technical mechanisms and standards that "fully protect privacy while taking advantage of the possibilities of digital technology to maximise the speed and real-time capability of national pandemic responses" (Pan-European Privacy-Preserving Proximity Tracing (PEPP-PT) 2020). The project promises the enforcement of data protection, including the anonymisation of data, in compliance with the GDPR and the requirements of IT security. Smartphones that run apps send out temporarily valid, "anonymous" identifiers (IDs), which are received and stored by other smartphones. The project promises that the contact history cannot be accessed by anyone, including the users themselves. Older events in the history would be deleted as soon as they became epidemiologically irrelevant. If users of the app are found to be infected, they will be contacted so that they can trigger the notification of their contacts. The project promises that only "anonymous" contact IDs from the contact history are transmitted to the server. Apps regularly download updates from the server, including the IDs of infected users' contacts, which allows users to find out that they have been in contact with infected people. Every app developed according to this standard must be certified by the PEPP-PT consortium (Schulzki-Haddouti 2020).

---

[3] The information can be found in the FAQ under Österreichisches Rotes Kreuz 2020b as well as in the data protection policy, Österreichisches Rotes Kreuz 2020a.

The project Decentralized Privacy-Preserving Proximity Tracing (DP-3T) intends to technically implement PEPP-PT in a – according to its own claims – decentralized form. Nevertheless, this system also requires a central server which only stores the necessary data and which therefore does not need to be trusted because "it does not maintain any secrets" (DP-3T-FAQ). First and foremost, the developers try to guarantee the secrecy of the data, identifying "secrecy of data" as the central starting point for data protection breaches.[4] To this end, "as much sensitive data [is stored] on user devices as possible" DP-3T Project 2020b. A risk score is calculated based on the locally available data as well as the data of infected persons, i.e. their (mathematically linked) IDs, which are regularly updated by the server. This score is supposed to reflect the probability of an infection (Troncoso et al. 2020). In addition, the app allows the transmission of "anonymous" data about all contact events that users had with all persons known to be infected for epidemiological research purposes (Troncoso et al. 2020).

The "Corona Data Donation" App of the Robert Koch Institute (RKI) is advertised as an offer to "support public authorities in coping with the most serious social crisis in 100 years" (Apple App Store 2020) and, according to the institute, is "used exclusively for scientific purposes" (Robert Koch-Institut 2020a). It serves neither the infected nor the uninfected, but solely the RKI.[5] It is intended to enable the RKI to automatically monitor persons who show symptoms that are typical for an infection with the corona virus, down to the postcode level, and to visually display the spread of the infection on the RKI website as a map updated daily. The use of the app requires the creation of a pseudonym of the user. Further data collection is partly done by manual input, partly by sensors such as wearables or fitness trackers that are worn on the body and connected to the app. A subset of the sensory data is anonymised, for example by mapping the weight to a weight range.

---

[4] This follows from the fact that all the risks described are formulated in such a way that they can only occur if the data becomes known.

[5] The term "data donation" was coined by the Deutscher Ethikrat (German Ethics Council) at the end of 2017 in order to grind down informed consent based on a narrowly defined purpose (on this and on the following Engel and Rötzer 2020). This framing was adopted by the Bertelsmann Foundation in 2018 in order to remove "structural hurdles" in data protection and to "meet personal reservations in a communicative way" (Bertelsmann Stiftung 2018), and in 2019 by the Federal Council of the CDU, which declared that it was about "moving away from data minimisation towards data sovereignty" (CDU-Digitalcharta 2019, p. 3). The construct of "voluntary data donation" serves, just like that of "data sovereignty", to lower data protection standards.

Linus Neumann has formulated a proposal for a contact data app (Neumann 2020) with reference to and using submissions to the hackathon "We vs. Virus". The sole purpose of this app is to enable a person who has been diagnosed with an infection to inform their contacts of the last 14 days. Neumann states that all data should be kept exclusively "decentralized" and "anonymous". To this end, anonymous IDs are generated locally on the smartphone at short intervals, which are sent out from the smartphone via Bluetooth Low Energy Beacon, which can then be received by other smartphones and, if the distance between the two is small enough, stores them. In the event of an infection, the smartphone sends its own anonymous IDs to a central server, which makes them available for retrieval by other smartphones, which can then determine locally whether they have had contact with infected persons.

Apple and Google partner on COVID-19 contact tracing technology (Google, Inc. 2020) that bears great similarities to DP-3T, is also decentralized, and only its own temporary IDs are transmitted to the server in case an infection is diagnosed. They plan to develop in two stages first an API and then a comprehensive, Bluetooth-based contact tracing platform built into their own operating systems.This API will allow for uploading one's own temporary IDs, which in case of infection may be signed by the public health department, to the server, or to compare the IDs downloaded from the server with the IDs one has seen oneself in order to determine the risk of exposure. It shall not be possible to extract the temporary IDs from one's own app.

## 2.1 Actors and constellations of actors

The social context in which the data processing procedures addressed here are carried out is characterised by a large number of actors who, against the backdrop of their mutual relations as well as the strategic and tactical interests they pursue, can and do influence the processing in various ways.

The EU Commission is currently trying to seize control over the conditions for the design and use of data processing methods for various purposes in the fight against the corona pandemic. On 9 April 2020 the Commission presented recommendations for "a common Union toolbox for the use of technology and data to combat and exit from the COVID-19 crisis, in particular concerning mobile applications and the use of anonymised mobility data" (European Commission 2020a), followed by "a European roadmap to lifting coronavirus containment measures" (European Commission 2020b) as well as the EU's eHealth Network's paper on "Mobile applications to support contact tracing in the EU's fight against COVID-19" () on 15 April 2020. In addition, the EU Commission promised a "Commission guidance on privacy and data protection", which is still outstanding.

German Federal and Laender Governments and Parliaments can set their own and the particular interests of individual actors as general interests and set them as binding law. As things stand at present, the German Federal Government is pursuing a multi-pronged approach: Health Minister Jens Spahn (CDU) recently wanted to amend the German Infection Protection Act in order to empower health authorities to identify possible contacts with infected persons by using mobile phone data that telecommunications providers should make available to the authorities. After much public criticism, the attempt was stopped, although not abandoned (Rudl 2020). In the meantime, the Federal Government massively supports the RKI's "Corona Data Donation" app.

The largest and most differentiated group of actors is the data subject. With regard to the data processing and its effects considered in this DPIA, from a fundamental rights perspective particular attention must be paid to the following groups of data subjects:

- infected data subjects,

- non-infected data subjects,

- data subjects who are de facto forced to use a corresponding app, for example by their employers or due to peer pressure,

- data subjects who have no or no compatible smartphone, such as typically children, mentally handicapped or older people,

- third parties that have a direct interest in the identification of infected persons as well as in accurate predictions regarding the spread of the corona virus.

Controller is a role within the scope of the GDPR. Controller is who "alone or jointly with others, determines the purposes and means of the processing of personal data" (Art. 4 No. 7 GDPR).

Among the actors in the technical field there are manufacturers, operators and service providers of various kinds. They may be controllers, processors, or third parties within the meaning of the GDPR.

Manufacturers can be formal organisations, i.e. legal persons or authorities, or groups of persons, each with members directly involved in the development of the app, mainly as software developers.

Operators of the server or the serves are generally formal organisations.

Service providers are active in different places, be it as operators of the data centers where the servers are located or the cloud systems on which the server runs as a service, as internet service providers who provide for data transport to and between the servers, or as mobile phone providers who provide for data transport to the smartphones. In addition, there are service providers who provide basic system functions, including the manufacturers of the smartphones and the smartphone operating systems, above all Google, Apple and Microsoft, who also operate their own central app stores, the providers of common frameworks or integration tools that are also used by government agencies, such as Palantir (Lewis, Conn, and Pegg 2020). Mobile phone providers such as Deutsche Telekom have voluntarily and without legal obligation "anonymized" mobile phone metadata of their users and transferred them to the RKI.

Security and law enforcement authorities, especially police forces, have extensive powers both to enforce pandemic containment measures, including curfews, and to access data held by public and private actors, both personal and anonymous, as well as IT systems. Furthermore, such data are expected to be transmitted to these authorities even after competent supervisory authorities have established the illegality of the transmission and ordered its cessation (Laufer 2020).

Supervisory authorities are independent authorities responsible for monitoring the application of the GDPR to ensure that the fundamental rights and freedoms of natural persons are protected when personal data is being processed. Their primary task is to monitor and enforce the application of the GDPR. In order to carry out its tasks, it has the necessary powers, for example to conduct investigations in the form of data protection reviews or to impose processing restrictions and prohibitions as well as fines.

Health authorities, which are police forces in the material sense, which are subject to competencies of the Federation or the Länder, have comprehensive powers with extensive rights to infringe on fundamental rights, in particular within the scope of application of the German Infection Protection Act (IfSG).

The Robert Koch Institute (RKI) is an independent German higher federal authority for, among other things, infectious diseases, which reports directly to the Federal Ministry of Health. In the area covered by the present DPIA, its task is to identify, prevent and combat the corona pandemic, to conduct related epidemiological studies and to report to the Federal Government as well as to the public.

Public and private health insurance companies and insurance companies collect, store, process and use large amounts of personal data, including special categories of personal data pursuant to Art. 9 GDPR, of insured persons and their relatives. There are a large number of legal regulations that oblige health insurances and insurance companies to transmit personal as well as anonymised data to third parties.

Healthcare institutions, hospitals, physicians and pharmacies collect, store, process and use large amounts of personal data, including special categories of personal data pursuant to Art. 9 GDPR, from patients. There are a large number of legal regulations that oblige these bodies to transmit personal as well as anonymised data, including data on infected persons and infections to health authorities under the IfSG.

Care institutions not only collect, store, process and use large amounts of personal data, including special categories of personal data according to Art. 9 GDPR, of persons in need of care and their relatives. They also exercise a great deal of control over the daily routine of persons in need of care. This also applies, albeit to a lesser extent, to outpatient care. They too are legally obliged to a large number of transmissions of personal as well as anonymous data.

Public research institutions are subject, with exceptions, to the sovereignty of the Laender, fulfil the tasks assigned to them under relevant Federal and Laender laws and ensure the freedom of research of their members. In recent decades, public research institutions have become increasingly dependent on the acquisition of third-party funding from public and private sponsors, including contract research, not least because of funding cuts or cost increases that the institutions have not been able to cope with. Private research institutions have to bear not only the individual research projects but also their institutional costs through their own or third-party funding. All research institutions are under increasing evaluation and assessment pressure on the one hand and time pressure with regard to publications on the other hand, the same applies to their members. Furthermore, the institutions are basically in competition with each other for personnel and money.

Operators of commercial Bluetooth tracking infrastructures, for example in shops, malls or on billboards in public spaces, track devices that are accessible via Bluetooth and their address data in order to deliver personalized advertising to the devices.

Employers, landlords and other entities with domiciliary rights control access to and use of facilities, buildings, premises or land, whether it be the workplace, the rented property, business premises, public transport or cultural institutions. They may enable, impede or prevent access and use.

## 2.2 Interests and constellations of interests

The different actors in the context of the processing procedure under consideration each pursue their own, overlapping or unrelated, shared or conflicting, strategic or tactical interests, which can influence the procedure positively or negatively, which are themselves influenced positively or negatively by the procedure, and which can make it either more plausible or less likely that these actors act as attackers in terms of data protection.

The main interest of those who use the app is that the app should enable them to determine whether they have been in (longer and closer) contact with infected persons and thus whether there is a risk of infection beyond a reasonable threshold. In addition, from the point of view of those affected, there is an interest that the app should enable them to prove to third parties that they have been exposed to a high risk and are therefore possibly infected in order to be medically tested for corona infection. At the same time, in order to prevent possible discrimination and stigmatisation, it is in the interest of the affected persons to be able to plausibly deny that they were exposed to a high risk, for example if the discrimination and stigmatisation take place exclusively on the basis of the risk index to which the affected persons were exposed and not on the basis of a medical diagnosis. It is also in the interest of the data subjects that the processing activity and/or the app enables them to prove that they were infected, but are no longer infected – and thus (probably) immune.

Since the processing of the data also affects persons who do not want to or cannot use the app, for example because they do not own a smartphone or do not have a compatible one, these persons have an interest in the processing procedure being designed in such a way that it can be carried out as far as possible without the use of an app, while offering the same or comparable functionality.

The main interest of government agencies, including the European Commission, Federal and Laender governments, and safety and health authorities, is to stop the corona pandemic or at least to control and manage it. In addition, these bodies have an interest in being able to track the spread of the pandemic in order to be able to make predictions about its further spread on this basis and to take appropriate measures accordingly. A third interest of the state authorities is the replacement of the general curfews by target group and risk specific arrangements, for which it is necessary to determine who is (probably) infected, for example because they have been in (longer and closer) contact with infected persons, and if necessary to inform these persons about their exposure. In addition, there is an interest of the state in being able to check and ensure compliance with the measures, whether they are curfews, restrictions on contact or quarantine requirements. And last but not least, there is the interest of the state, which is clearly evident in public announcements by political decision-makers, in the tightening of laws, and most recently also in violations of orders prohibiting transmission (Krempl 2020), in the expansion of surveillance infrastructures, including the necessary legal foundations for this, also for future ones, as yet undefined purposes, a perpetuation of existing surveillance measures and their increasing extension to ever broader purposes. The state's lack of interest in effective protection of fundamental rights is demonstrated not least by the fact that neither the EU Commission, nor the Federal or Laender governments, nor the European, Federal or Laender legislatures have ever conducted a DPIA pursuant to Art. 35 (10) GDPR in connection with the enactment of legislation governing the processing of personal data, even though the GDPR was adopted four years ago. This was not done even in cases where there was clearly a high risk to the rights and freedoms of natural persons.

Supervisory authorities notoriously suffer from a lack of resources, both in terms of staff and material resources and, in particular, in terms of the technical skills of their staff. They therefore have an interest in efficient examination procedures. In the case of a high risk associated with the processing, they therefore expect the existence of a substantially meaningful DPIA, in which all relevant requirements determined from the GDPR are gathered together in a methodically comprehensible manner and can be presented and assessed in a way that is both functionally and normatively auditable.

Both public and private research institutions pursue the interest to understand the spread of the corona pandemic in order to be able to make predictions about its further spread, but also for general knowledge about pandemics, the effect of government measures on their spread and containment, the behaviour of persons in such pandemics, including their movements, and other phenomena in this field.

Manufacturers of the corona app – or in general: of all corona apps, especially if they are published as open source software – have an interest in selling the app or the app being used, while offering additional "comfort functions" and functions with other purposes, possibly on the basis of an additional informed consent, which could undermine the data protection rights of the users.

As economic actors, operators and service providers have an interest in being able to offer their services at the lowest possible cost for the highest possible price. To the extent that they can reasonably expect not to be audited or not to be audited comprehensively by the supervisory authorities, they have an interest in saving especially on those parts of their services that cannot be audited or are practically not or rarely audited. Instead of measures that stand in the way of a possible further or secondary use of systems and personal data, they will tend to concentrate on measures that primarily serve to safeguard the processing, but which can be sold to outsiders as data protection measures, such as data protection policies or the security measures pursuant to Art. 32 GDPR. Operators and service providers have an interest in the continuity of processing activities enforced by the state, because in this case they can reasonably expect to be selected for further operation as the ones who have already provided their service in the phase which was planned as being limited in time, if the processing activities are made permanent.[6] This includes in particular service providers who operate, control and provide basic system components or functions or common interfaces to other systems or platforms used either by the state or by private parties, such as Google, Apple, Facebook, Amazon or Microsoft, but also manufacturers or operators of third party systems, who wish to use such systems, use such interfaces or further process the personal data processed in the processing activity, such as Palantir.

Employers, landlords and holders of domiciliary rights have an interest in being involved in the processing activity as receiving parties in order to be able to decide, on the basis of the information obtained, whether or not to grant access to buildings, premises or other facilities, either for the purpose of minimising as far as possible the risk of infection to employees, customers or guests, or for reasons of personnel management and control, thereby externalising any risks to the data subjects.

Third parties, be they individuals, groups, or governmental or private organisations, have a variety of different interests. In particular, the following interests are relevant for the processing activity examined here:

– the interest in revealing the identity of infected persons, whether for personal, economic or state reasons,

– the interest in non-valid data on the spread of the pandemic, whether the case numbers are too high or too low,

– the interest in disrupting the processing activity, whether for the pleasure of "hacking", for the pleasure of destruction or with the aim of undermining public

---

[6]Cf. the cooperation announced by Apple and Google on 10 April 2020 (Google, Inc. 2020).

trust in the ability of the State or controllers to operate such a processing activity successfully, securely and data protection compliant,

– the interest in exploiting particular characteristics of the processing activity, such as the permanently enabled Bluetooth device that is necessary for steadily broadcasting TempIDs, the upload of Health-tempID, or the Download of non-identifying infection-indicating data (NIID) for their own purposes, such as to attack the smartphones that have their Bluetooth enabled or the servers involved.

# Chapter 3

# Use cases

The design of the data processing procedure, in which the corona app is embedded, is based on fundamentally different assumptions about the typical use of the App by the users and of the overall system by the operators. In the following we reconstruct these assumptions as use cases. Their breakdown helps to understand and verify the guiding ideas behind various design features of the app, the server and the processing activity as a whole.

## 3.1 The processing procedure

(1.1) The processing procedure, in which the app is embedded, allows and enables interventions in case of failures.

(1.2) The procedure provides for the revocation of invalid, incorrect or erroneously transmitted Health-TempIDs or the corresponding non-identifying infection-indicating data (NIID), for example if the diagnosis of the infection turns out to be incorrect.

## 3.2 Legal bases and compliance

(2.1) There are valid, constitutionally compliant and fundamental rights-friendly legal bases for the processing activity according to the definition in Art. 4 (2) GDPR, which also regulate the accountability and responsibilities.

(2.2) There are corresponding legal bases for the production, provision and operation of the server(s) with which the apps communicate and which serve to receive health TempIDs from infected persons, to anonymise them and to distribute the non-identifying infection-indicating data (NIID).

(2.3) There are corresponding legal bases for the production, provision and maintanance of the app.

(2.4) The rights of the data subject (Art. 12–22 GDPR) are safeguarded.

(2.5) The controller as well as the processor within the meaning of the GDPR behave at all times in accordance with the law.

(2.6) The controller and the processors ensure that their personnel behaves at all times in accordance with the law.

## 3.3 Operation of the technology

(3.1) The server or servers, including the underlying infrastructure, are set up and operated securely (confidential, with integrity and available) and data protection compliant (transparent, non-linkable and intervenable) according to the state of the art.

(3.2) The server or servers are available at all times.

The smartphone

(4.1) The person owns a smartphone. She downloads the app from a secure source, perhaps from the operator's website or the app store operated by her smartphone's operating system manufacturer, and installs the app according to the installation instructions. After its installation, the app is operational.

(4.2) The person carries her smartphone with her at all time when she is near other people. This also applies within the family environment.

## 3.4 The App

(5.1) The app is being developed using state of the art methodologies as well as implemented securely (confidential, with integrity and available) and data protection compliant (transparent, non-linkable and intervenable).

(5.2) The app transmits changing IDs ("TempIDs") via Bluetooth Low Energy Beacon ("TempID token") on a regular basis, approximately every 5, 15 or 30 minutes, and receives the TempIDs of other smartphone apps in the environment. The validity period of the TempIDs is set by the manufacturer in accordance with the controller. The measurement only yields valid data that is stored correctly in the app.

(5.3) The app regularly contacts the server(s) and downloads updates with the non-identifying infection-indicating data (NIID). This data is generated on the server(s) by anonymising Health-TempIDs submitted to the server(s) by persons diagnosed as infected. The downloaded data is correct and valid, it is transferred without errors and is stored correctly in the app. Communication with the server(s) is confidential and repudiable towards third parties. The communication with the server(s) is not linkable to other communications with the server(s).

(5.4) Based on local calculations, the app determines whether exposure has occurred using the contact data stored in the app, i.e. third-party TempIDs stored due to contact events, the time duration and signal strength profile, and the non-identifying infection-indicating data (NIID) received from the server or servers. The calculation method, according to which exposure is calculated on the basis of contact events, is carried out according to the specifications of the competent health authorities and produces valid exposure risks. The calculation procedure is transparent and public, and is implemented in the app without errors.

(5.5) Warnings may be accompanied by further information, such as additional sources of information, or instructions for action. The instructions for action are defined by law, are public and transparent, and are correctly implemented in the app.

(5.6) The TempIDs generated by the app will be deleted from the app after the time periods specified by the competent health authorities.

(5.7) The third-party TempIDs of contact events stored in the app will be deleted from the app after the periods specified by the competent health authorities.

(5.8) The revoked non-identifying infection-indicating data (NIID) are deleted in the app immediately after the update.

(5.9) As soon as a positive match has been calculated, and thus the purpose of the app has been fulfilled, all TempIDs as well as the TempID history in the app will be deleted. A message is displayed indicating that the app can be uninstalled.

(5.10) The transmission of Health-TempIDs from the app to the server(s) is logged in the app with integrity. The transmission log is both provable and repudiable to third parties.

(5.11) The TempIDs used in the app are generated according to a state of the art procedure so that they are unique worldwide.

(5.12) The app is available for download at any time from the manufacturer, the controller and in the app stores.

(5.13) The app is available on the smartphone at all times after installation.

(5.14) The TempIDs are avalailabe at all times, both in the app and on the server(s).

## 3.5 The Person

(6.1) If the person has been diagnosed with a coronavirus infection, the person will use the app to transmit the TempIDs that the app has generated and broadcasted within the last few days as Health-TempIDs to the server(s). The number of days prior to the time of diagnosis for which the Health-TempIDs are sent is determined by the manufacturer according to the specifications of the responsible health authorities. The default number of days is public and transparent, and is correctly implemented in the app. The transmission of the Health-TempIDs to the server must be actively triggered by the person.

(6.2) After an infection the affected person goes into (self-)quarantine and deactivates the app to prevent false negatives.

(6.3) As long as a person has not been diagnosed with an infection, neither she nor the app will transmit data to the server(s).

(6.4) If the risk of exposure exceeds one or more pre-defined thresholds, the user will receive exposure alerts. The threshold(s) are specified by the competent health authorities. The user may also be able to specify an additional threshold of her own. The thresholds are transparent and public. All threshold values are correctly stored in the app.

(6.5) The app allows the user to effectively repudiate her own exposure, which does not indicate an infection, to third parties, for example to reduce the risk of discrimination or stigmatisation.

# Chapter 4

# Description of the processing activity

The controller must provide a description of a processing activity for the implementation of a DPIA. So far, only drafts for processing with the help of a tracing app are available. In view of the unclear political and legal situation, it is not to be expected that a sufficiently robust description of the processing activity will be provided by the controller. Therefore, this document refers to a specified processing operation based on the drafts of DP-3T and Linus Neumann. This specification is presented in "high-resolution" in the chapter 3 "Use Cases", showing legal and functional assumptions.

A description of the processing must make statements about which functional, security-related and data protection-friendly properties are being aimed for or what has already been implemented; i.e. for example, which concrete protective measures have been planned or implemented in detail. This is not sufficiently possible in this case, which is based on drafts. Instead, only the protective measures to be taken can be specified here. In the deeper technical description layers, further, technically more detailed possibilities are discussed.

According to Art. 35 GDPR, the responsible party is obliged to prove the effectiveness of its measures before the processing is put into operation. The processing of the data must be permanently verifiable (cf. Art. 32 para. 1 lit. d GDPR). This means that a data protection management (DPM) must be in place for operations, which not only controls, checks and assesses the requirements of the GDPR, but also effectively enforces them in the organisations involved. The testability of a processing is not a passive characteristic, but must be actively established during the design phase of the processing activity (Seidel 1984, p. 191). Logging is an essential prerequisite to realize testability of the activities of persons and IT systems in the context of processing. In this case of high-risk processing (see chapter 6), revision-proof, i.e. integrity-based, logging is necessary, especially at those points in the processing when there is an information relating to an identified or identifiable natural person is to be removed. This means that at least the following aspects must be part of a description of the processing activity:

1. The nature, scope and circumstances of the processing activity (see chapter 4.1).

2. Generally understandable description of the purpose and functionality of the processing activity with which this purpose is to be achieved (see chapter 4.2).

3. Description of assumptions, as to why the purpose is legitimate (see chapter 4.3).

4. Description of foreseeable "neighbouring" purposes (see chapter 4.4).

5. Identification of the categories of personal data used (see chapter 4.5).

6. Analysis of the individual processing activities (see chapter 4.6).

7. Identification of the measures specifically designed to ensure that the purpose is limited and to make it verifiable (see chapter 4.7).

8. Designation of protective measures required in particular under Art. 5 GDPR (principles), Art. 25 GDPR (Data protection by design and by default) and Art. 32 GDPR (security) (see chapter 4.8).

9. Identification of further requirements of the GDPR (such as legal basis, Data Protection Officer, Data Protection Management) (see chapter 4.9).

10. Designation of the controller (see chapter 4.10).

These aspects are discussed below.

## 4.1 Nature, scope and context

This type of data processing is at least new in the sense that an epidemiological problem is solved by a procedure which is supposed to be based on a broad use of networked information technology systems, the cooperation of many people, and the use of data with requiring a high level of protection (medical data), is instructed. Although the planned individual technical components for the support of the procedure and the combination of these components is rather a novel approach, without practical experience so far.

In any case, data processing is extensive, since it processes personal information of a very large number of people on the scale of entire states, continents or even the entire planet.

The circumstances of this data processing are, if not unique in their occurrence, exceptional. As a rule, most data processing activities have predecessors and variations that have been introduced, operated, and decommissioned in other organisations. This is not the case here.

## 4.2 Purpose of processing

The overall purpose of the processing of personal data is to control or contain the global corona pandemic, from which the detection and interruption of infection chains is derived. This purpose specifically includes the rapid information of potentially infected persons. In order to achieve this purpose, a reliable indicator is to be made available to those affected if they have previously had epidemiologically relevant contact with people infected with SARS-CoV-2. Thereupon, the warned persons are to be placed in home quarantine to interrupt the chain of infection. The incubation period of COVID-19 is up to two weeks and infected persons are already contagious before possible symptoms of the disease appear. This makes it necessary to record a contact history in order to identify the contacts retrospectively. The duration of the contact history is specified as a parameter by the health authorities.

The procedure includes the following processing activities:

### a) App-side processing of contact events

In the best case scenario, an app (corona app, CA) is installed on all smartphones used by citizens, which uses Bluetooth technology (Bluetooth Low Energy Beacon) to send out a random string of characters – a so-called temporary identification feature (TempID) – at regular intervals as a TempID token. This TempID changes regularly within a day, for example every 5, 10, 15, or 30 minutes. If smartphone A with an installed app receives a signal of a certain strength from smartphone B (also with app), the distance between the smartphones is estimated. The stronger the signal between the smartphones, the smaller the distance between the people. If the distance between the persons is small enough, both apps unpack the respective other TempID token and save the unpacked, foreign TempID.

The own TempID and the received foreign TempIDs sent in this way are stored in the own app for a limited time are personalized for this time. They are person-related because they are stored on a smartphone (in the app) that is assigned to a person. Since the TempIDs sent out will change over time, users cannot track each other permanently in everyday situations.

### b) Authorisation of uploading, anonymisation, temporary storage and dissemination of positive infection status

Assuming smartphone user A is later diagnosed as infected. In this case, she will receive a TAN, whether from a doctor or an authority, with which she can authenticate the upload of TempIDs, which became Health TempIDs through the diagnosis, to the CA server. After the upload, the data is anonymised in a legally, organisationally, and technically effective manner, and the non-identifying infection indicating data (NIID) is stored on the server as a central buffer. They are then distributed via updates to all apps used.

### c) Decentralized contact tracking

All other smartphone users regularly download the published non-identifying infection indicating data (NIID) into their app. These TempIDs are without any statement about the connection to an identified or identifiable natural person (Article 29 Data Protection Working Party 2007). At the same time, they only have an informational value for all those persons who had contact with the infected persons within a certain risk area.

The comparison of the non-identifying infection indicating data (NIID) published on the server with the TempIDs stored locally on the smartphones makes it possible to determine whether this user had contact with a presumably contagious person during a certain period of time. However, it cannot be deduced from the data whether the contact was with a single infected person over a longer period of time, or whether it was rather several short contact events with many different people. Such details cannot be reconstructed from the data. Using a programmed calculation rule, a risk score is calculated from the contact events.

There is no data on the central server about which people are infected, where they have been and when, or which people they have met where.

The process by which a one-time TAN, unsable once and only by the person tested positive, can be generated and transmitted to the diagnosing doctors. A TAN can play a special role, when it is exchanged with its first half among persons presents, namely patient and doctor, while the other half of the TAN is communicated to the patient by telephone after the result of the laboratory test. By entering the two TAN parts, the upload to the server is initiated so the Health-TempIDs are uploaded to the CA server. If the upload is initiated during the telephone call, the doctor can deactivate the validity of the TAN after the end of the telephone call, otherwise the patient must call the doctor again after the upload to deactivate the TAN.

As soon as a positive match is detected, i.e. there is a risk of infection, all TempIDs in the app can be deleted or the app can be uninstalled. The time period that must be taken into account in order to delete already uploaded non-identifying infection indicating data (NIID) on the CA server, e.g. in the case of a non-existent infection, must also be taken into account.

## 4.3 Legitimacy of the purpose

The interruption of chains of infection by warning people who may be infected, with reference to a quarantine recommendation, is not only legitimate but also necessary in the case of a highly contagious, and in the worst case, fatal disease.

The processing activity used for this purpose and the techniques used, such as an app with a server, must meet the requirements of data protection and IT security. Processing activities that are particularly carry the burden of proof in terms of fundamental rights that they fulfil the purpose which was defined as narrowly as possible, and that there is no milder variant that protects fundamental rights (cf. 5.4).

## 4.4 Distinction from "neighbouring" purposes

In the discussions about the corona app (CA), different expectations are placed on the functions of such an app. As a rule, attention is focused solely on the CA and not on the processing activity as a whole, of which the use of the CA is only a part.

Pursuing purposes other than those mentioned above means that they are different processing operations that have to undergo a different threshold and data protection impact assessment (DPIA) than the one presented here, which in turn may reveal different risks and different safeguards. This DPIA is limited solely to the processing for the above-mentioned purpose of identifying and interrupting possible chains of infection by informing potentially CV-infected persons.

Other purposes – and thus other functions of the CA – would be for example (a) the tracing of the epidemiological spread of CV, (b) the warning of CV-infected smartphone owners in spontaneous encounters, (c) the surveillance of CV-infected persons, (d) the establishment of forecasts of epidemiological spread, (e) the treatment of CV-infected persons who own a smartphone.

### I. Tracking the epidemiological spread of CV

From an epidemiological point of view, it is desirable and legitimate to be able to trace the geographical distribution of CVs. This functionality has so far been guaranteed by report of the health authorities.

With CA type 3 decentralised, a presumably sufficiently resolved localisation of the CV would be possible if there were not just a single central server but many geographically distributed servers and infected persons would be required to report to the nearest server, e.g. the one of the responsible health authority. Otherwise, a processing activity based on CA type 3 would only be able to decentrally determine the fact of reported cases, or their increase or decrease, provided that infected persons with CA type 3 are obliged to upload their health TempIDs decentrally. This would be a secondary purpose which a CA type 3 could fulfil decentrally and would, if made part of the original purpose, entail an extension of the DPIA to identify further risks.

## II. Warning of CV-infected smartphone owners during an encounter

From the direct point of view of the data subject, it would be desirable and, in principle, legitimate for holders of a CA to automatically inform each other about their status regarding a CV infection by means of an app during everyday spontaneous encounters. However, this purpose entails processing that works intensively with personal data.

On the other hand, the benefit of such an app to prevent contact with CV-infected persons would be small, as it is to be expected that the identified CV-infected persons will no longer be in the public eye or will deliberately conceal their infection.

The benefit of such an app with persons who have not yet been diagnosed as infected when tested would also be small, because the incubation period is variable, from a few to 14 days, and there is a risk of infection even in the absence of symptoms.

The use of an app that allows tested and immunized persons to report their status via Bluetooth could be considered, but is not within the scope of this DPIA. This purpose is also not realisable decentrally as a secondary purpose of the CA Type 3.

## III. Monitoring of CV-infected persons

One of the most frequently formulated requirements of a CA is to be able to monitor CV-infected persons or possible candidates via app.

In this performance, an app would have the function of an electronic anklet. CV-infected persons would be obliged to make any movement outside their quarantine quarters – or inside a hospital – only by taking their smartphones with them. There would then have to be an instance that monitors such movements and, if necessary, triggers an alarm in case of risky encounters, either at a reporting office, or directly for other smartphone users. Alternatively, risky encounters would have to be recorded in order to be able to prove to CV-infected persons afterwards that they did not comply with the regulations.

In the "Corona Data Donation" app published by the RKI on 7 April 2020, the function of monitoring the health data of individuals is realised by linking a smartphone app with fitness trackers, so that together with the localisation of users on the basis of postcodes, there is also a locally specified full monitoring of the bodily functions of users. However, such an application scenario also pursues a completely different purpose while representing a completely different processing activity. This purpose could also not be implemented decentrally as a secondary purpose with a CA type3 and would entail a much more intensive infringement on basic rights.

## IV. Creation of epidemiological distribution forecasts

The purpose of making predictions about the epidemiological dissemination goes one step further than that described in (a). Scientific forecasts attempt to make statistical statements about the future situation, to determine for example the effect, of certain measures or the imminent need for resources (hospital beds, drugs).

In order to implement this purpose, data on persons already diagnosed as infected and their approximate geographical position would probably be necessary in order to keep restrictive measures and additional resource expenditure as minimal as possible and regionally limited. For the preparation of forecasts, this data would have to be stored centrally and evaluated in a methodically controlled manner.

In the app published by the RKI on 7 April 2020, this function is implemented by a smartphone app that localises people on the basis of a postcode entered by the user (Robert Koch-Institut 2020b. Als Zweck wird ausgewiesen:)

> "Making the data from my fitness bracelet available will help the Robert Koch Institute (hereinafter referred to as "RKI") to better predict the course of the disease with COVID-19 nationwide and thus improve the control of containment measures against the corona pandemic. The predictions are to be made on a daily basis at postcode or district level and made available to the public in anonymised form. On the basis of scientific models, the app uses my personal data to calculate the probability of the presence of a flu-like illness such as Covid-19 on a daily basis. Even the evaluation of the resting pulse, sleep duration and activity level are sufficient for the detection of corresponding symptoms. My individual data will be merged with the data of all other app users and evaluated nationally / regionally (in the following "purpose")."(Robert Koch-Institut 2020b)

This allows the epidemiological spread to be predicted by monitoring bodily functions of persons whose address data are accessible to the RKI.

More recently, such projects have been supported with techniques from the field of "big data" and so-called "artificial intelligence". The risks associated with an IT-supported statistical analysis of this scope, in which entire populations are evaluated, have most recently become apparent in the Cambridge Analytica case. In this case, the data of millions of Facebook users was used for the purpose of manipulating democratic elections. The larger the centrally managed amount of data, the greater the desirability and risk of abuse.

Although this purpose could be added to (a), it would have new serious consequences and risks for individuals, groups, and the society as a whole due to centralised storage and additional statistical analysis thus would therefore require a separate DPIA.

### V. Treatment of CV-infected smartphone owners

On April 8, 2020, the EU Commission presented a recommendation for a uniform Europe-wide standard. In recital 13, it identifies among other things the treatment of CV patients with the aid of medical devices as a legitimate purpose (European Commission 2020a). This could also include the CA.

This purpose is non-specific and may include the possibility to give COVID-19 patients only recommendations for self-treatment via the CA, to establish a text-, audio- or video-based communication channel between doctors and patients or for the imposition of coercive measures on COVID-19 patients by an authority.

There is a lack of a precise definition of purpose and the processing operation, which can serve as a basis for an independent DPIA can be carried out.

## 4.5 Categories of personal data used

Categories of personal data used, the recipients of such data and, where appropriate, a transfer to a third country or an international organisation, are described below.

The transfer to a third country is not excluded in the context of cloud computing. The cloud services of large providers may be located outside the EU.

Transmission to international organisations could also take place within the framework of transnational cooperation to combat the pandemic.

**Temporary pseudonymous identifier (TempID)** In the information model, the temporary pseudonymous identifier (TempID) maps the temporally limited existence of a CA user in the vicinity in which infections can occur. In itself, this TempID does not contain any information about place, time or person, but it is personal, as long as it is located on the smartphone of a person, the sending or the receiving one. This information is changed at regular intervals, for example every 5, 10, 15 or 30 minutes. The recipients of this information are other CA users within Bluetooth range.

**Distance** The distance is the estimated distance between two smartphones within physical range. It is determined by sensors using the Bluetooth signal strength of the received TempIDs. However, the constructed relationship between signal strength and distance is based on different assumptions about the physical and technical situation. For example, one assumption is that the location of the smartphone is identical to the location of the CA usersïn in three-dimensional space. Another assumption is that there are no space-dividing elements in the contact field (for example glass walls).

The respective data is stored together with a received foreign TempID.

**Duration** The duration refers to the registered number of identical TempIDs in relation to time. This date is stored on the smartphone of the CA user together with the received foreign TempID.

**Infection status** The infection status has several CV states: Not infected, exposed, infected, immune. The first state is the default state. The second state results from contact events with infected persons. The third and fourth states depend on a medical diagnosis.

In the case of the infection status "infected", this information is communicated to the server by transmitting the Health TempID. After anonymisation, only non-identifying infection indicating data (NIID) is on the server.

**Risk-Score** If there are matches between the sensorically received TempIDs and the non-identifying infection indicating data (NIID) received from the server, an individual risk score is calculated on the basis of the duration and distance profile of the contact events with the aid of the statistical calculation rule. If this score is above a defined threshold value, notification of the user is triggered.

**IP addresses** The IP address is necessary for communication on the Internet (routing, addressing). Due to the Internet architecture, the recipients of this data are primarily the server operators and secondarily the Internet providers.

The IP address is required as a minimum in three different application cases: (a) when retrieving an app from an installation source on the Internet, (b) when retrieving non-identifying infection indicating data (NIID) from the server, and (c) when reporting one's own health TempIDs to the server if the user has been diagnosed with an infection.

**Time specification** The time specification indicates the date of the contact. This date is stored together with the received foreign TempID on the smartphone of the CA user.

**TAN** The TAN is generated by the controller or a trustworthy third party and transmitted by the treating physician to a person diagnosed as infected. This TAN regulates access control on the server and allows the person, and only the person, to transmit their health TempIDs to the server.

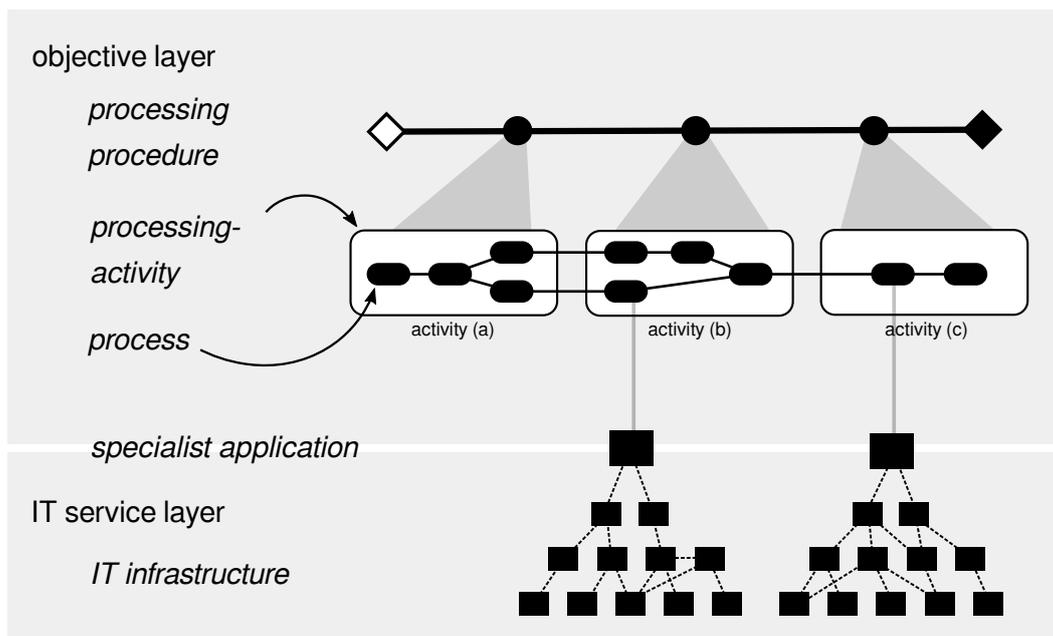

Figure 4.1: Structure of the processing procedure

## 4.6 Analysis of the individual processing activities

In this section, the processing activities are successively broken down into so-called operations under Art. 4 No. 2 GDPR, which must be fair and manageable by the controller. Some of these processes are IT-supported and are considered in more detail in the analysis. A schematic overview is provided in the figure 4.1.

In the first step, the following three levels of analysis are considered for each processing activity:

1. **Object or purpose level,**
2. **Professionial (application) level,**
3. **IT service level.**

In the second step, the operational components physically required for the processing activity are described at all levels, namely:

– the personal **data** processed therein and, if applicable, technical, organizational and personnel **roles**,

– the technical **systems, services** and related operational **processes** involved

– the existing **communication relations** and the **interfaces** used for it.

The first (material) level has already been described in a general way in section 4.2 and is now examined here with regard to the processes or series of processes taking place there.

**Processing activity "App-side processing of contact events"**

**Material level:**

| No. | Procedure | Roles | Data |
|---|---|---|---|
| 0 | Installation and commissioning of CA | CA user | Usage data, device data |
| 1 | Local generation of own TempIDs | | TempIDs |
| 2 | Saving own TempIDs | | TempIDs |
| 3 | Transmission of own TempIDs | | TempIDs |
| 4 | Receiving foreign TempID tokens via BTLE | | Foreign TempID tokens |
| 5 | Save contact events | | Foreign TempIDs, distance, duration, time |

**Application level:**

| No. | Process | Specialist application(s) |
|---|---|---|
| 1 | Local generation of own TempIDs | CA |
| 2 | Saving own TempIDs | CA |
| 3 | Transmission of own TempIDs | CA |
| 4 | Receiving foreign TempID tokens via BTLE | CA |
| 5 | Save contact events | CA |

**IT services level:**

| No. | business application | IT systems / Services | Processes | Role |
|---|---|---|---|---|
| 1 | CA | Smartphone, BTLE service | Life cycle maintenance | CA users |

**Communication relations and interfaces:**

| No. | Source | Destination | Interface | Purpose |
|-----|--------|-------------|-----------|---------|
| 1 | CA1 | CA2 | Bluetooth | Transmission of TempID token from CA1 |
| 2 | CA2 | CA1 | Bluetooth | Transmission of TempID token from CA2 |

**Processing activity "Authorisation of upload, anonymisation, temporary storage and dissemination of positive infection status"**

**Material level:**

| No. | Procedure | Roles | Data |
|-----|-----------|-------|------|
| 1 | Medical examination (sampling) | doctor, CA user | TAN |
| 2 | Survey of the infection status (test procedure) | Medical staffl | |
| 3 | Authorisation of the assigned TAN and transmission to CA users | Doctors, CA users | TAN |
| 4 | Authorized transmission of the health TempIDs of CA users to the server | CA users | TAN, health TempIDs |
| 5 | Anonymization of health TempIDs on the server | Operators | health TempIDs, non-identifying infection indicating data (NIID) |
| 6 | Storage of the NIIDs on the server | operators | NIIDs |
| 7 | Deletion of NIIDs on the server 14 days after the epidemiologically determined time | Operator | NIIDs |

*4.6 Analysis of the individual processing activities*

**Application level:**

| No. | Process | Specialist application(s) |
| --- | --- | --- |
| 1 | Medical examination (sample collection) | TAN management |
| 2 | Survey of infection status (test procedure) | *unknown* |
| 3 | Authorisation of the assigned TAN and transmission to CA users | TAN management |
| 4 | Authorized transmission of the health TempID to the server | CA |
| 5 | Anonymization of Health TempIDs on the server | CA server |
| 6 | Storage of the non-identifying infection indicating data (NIID) on the server | CA server |
| 7 | Delete the NIIDs on the server 14 days after the epidemiologically determined time | CA server |

**IT service level:**

| No. | Business application | IT systems / services | Processes | Roles |
| --- | --- | --- | --- | --- |
| 1 | TAN administration | *unknown* | *unknown* | Doctor |
| 2 | CA | Smartphone, Internet | Life cycle maintenance | CA user |
| 3 | CA server, database server, server-OS, HW | RZ, internet | Server administration, RZ operation | Server admin |

*Chapter 4 Description of the processing activity*

**Communication relations and interfaces:**

| No. | Source | Destination | Interface | Purpose |
| --- | --- | --- | --- | --- |
| 1 | CA | CA Server | TCP/UDP | Health TempID transmission |
| 2 | CA server | CA | TCP/UDP | Retrieval of the list of non-identifying infection indicating data (NIID) |

**Processing activity "Decentralised contact tracing"**

**Sachliche Ebene:**

| No. | Procedure | Roles | Data |
| --- | --- | --- | --- |
| 1 | Queries of the list of non-identifying infection indicating data (NIID) from the server by other CA users | | List of NIIDs |
| 2 | Comparison of the non-identifying infection indicating data (NIID) with the TempIDs received via Bluetooth on the smartphone | | Lists of TempIDs and NIIDs, matches |
| 3 | Use the matches to calculate the risk of infection and the presentation of behavioural recommendations do nothing, quarantine, test | | Matches |
| 4 | Use the recommended actions to contact medical staff or the public health department if necessary | CA users | Documents, text, image, interaction |

**Application level:**

| No. | Process | Specialist Application(s) |
|---|---|---|
| 1 | Query of the list of non-identifying infection indicating data (NIID) from the server by other CA users | CA, CA server |
| 2 | Comparison of the non-identifying infection indicating data (NIID) with the TempIDs received via Bluetooth on the smartphone | CA |
| 3 | Use the matches to calculate the risk of infection and the presentation of behavioural recommendations (do nothing, quarantine, get tested) | CA |
| 4 | Use the recommendations for action to contact medical staff or the public health department if necessary | CA |

**IT-Service level:**

| No. | Business application | IT systems / services | Processes | Roles |
|---|---|---|---|---|
| 1 | CA | Smartphone, Internet | Life-Cycle-Management | CA-User |
| 2 | CA server | Database server, data center, internet | Server administration | Server admin |

**Communication levels and interfaces:**

| No. | Source | Destination | Interface | Purpose |
|---|---|---|---|---|
| 1 | CA server | CA | TCP/UDP | Transmission of the non-identifying infection indicating data (NIID) to the CA for local matching |

## 4.7 Designation of measures to limit the purpose

Essential measures to ensure the limitation of a processing activity generally consist of using pseudonymised and anonymised data, in which the information relating to an identified or identifiable natural person is as far as possible removed or made subject to conditions, and of separating the data files, communication relations, and sub-processes of this processing activity from other processing activities.

For the present procedure, this means that the TempIDs are not speaking, but ideally are globally unique random numbers. Without conditional contexts, they should not make any direct statements about persons, places, times, and social circumstances possible.

To enforce the limitation of the purpose on the smartphone, it is necessary to run the app in a container. The container should protect against access by other apps or by the operating system. In addition, the interface to the TempID tokens, as well as the download of the non-identifying-infection indicating data (NIID) and the upload of the health TempID to the server must be subject to special protection through authentication and encryption. These events must also be logged.

A very important information relating to an identified or identifiable natural person consists in this processing activity at the time when the physician transmits a TAN authorizing the upload of the health TempIDs to the server.The link created by the doctor between the TAN and the sick person is strictly confidential and must be protected by appropriate security measures, such as admission-, entry-, and access controls, authorization management, backups, encryption, integrity safeguards, deletion, and logging, including a mature data protection management.

A further measure regarding purpose limitation and thus unkinkability is the automatic deletion after the number of which is the competent health authority has defined as maximum contagiousness.

## 4.8 Designation of further planned protective measures

All processing activities must meet the requirements of the GDPR, as required in concentrated form by Art. 5 GDPR with the principles. In particular, the requirements of Art. 25 GDPR on data protection by design and default and of Art. 32 GDPR regarding the security and resilience (Gonscherowski, Hansen, and Rost 2018) of the processing are added.

The following protective measures were taken from the design proposals – especially from the design documents of the DP-3T project (DP-3T Project 2020a) – for the described procedural purpose and arranged according to the protective objectives of the standard data protection model (DSK SDM2.0a, Rost 2018a). The vulnerabilities and security holes that result in the protection goals being compromised or violated are described in section 7. The protection goal of unlinkability has already been addressed in section 4.7.

### a) App-side processing of contact events

This processing activity is technically supported by the use of a smartphone and the CA installed on it.

No additional technical or organizational measures for operational security are provided for the smartphone other than the general recommendations for the use of smartphones.

**Confidentiality** – The contact history is protected against unauthorized access by encrypted storage on the smartphone.

**Availability** – The app does not need to be unlocked, or otherwise personally activated, so that several smartphones can be used in parallel.

**Integrity** – None.

**Intervenability** – The participation can be terminated at any time by uninstalling the CA, switching off the Bluetooth module or the smartphone.

**Transparency** – This process step and the function of the CA is described in the process documentation.

- The data subject is informed about the data processing procedure by a data protection declarationin the app. Specification, documentation and source code of the CA are openly available, so the CA is testable.

**Unlinkability** – See section 4.7.

## b) Authorisation of uploading, anonymisation, temporary storage, and dissemination of positive infection status

**Confidentiality** – The data transfer between server and CA is protected against unauthorized access by using an encrypted and integrity-safe channel.

- TempIDs are used as pseudonyms,[1] which makes it difficult for the server operator to identify infected persons directly.
- The link created by the doctor between the TAN and the sick person must be kept strictly confidential.
- The server does not create any protocols (IP address or similar) of the upload processes.

**Integrity** – The data transfer between server and CA is protected against unauthorized access by using an encrypted and integrity-safe channel.

- Release of data transfer via TAN, which is only assigned by female doctors after a positive test.

**Availability** – None.

**Intervenability** – CA users initiate the notification of the infection status themselves. Release of the data transfer explicitly only after entering a TAN, which also increases the perception of relevance.

**Transparency** – This procedure step and the function of the server are described in the documentation of the processing operation. Specification, documentation, and source code of the app and the server are open, so the CA and server can be checked.

**Unlinkability** – See section 4.7.

---

[1] The DP-3T white paper, for example, erroneously speaks of anonymous IDs.

**c) Decentralized contact tracing**

**Confidentiality** – The data transfer between server and CA is protected against unauthorized access by using an encrypted and integrity-safe channel.

**Integrity** – The data transfer between server and CA is protected against unauthorized access by using an encrypted and integrity-safe channel.

**Availability** – Due to the central storage on a server, the list of non-identifying infection indicating data (iDoP) can be restored by re-downloading.

**Intervenability** – The data transfer requires an active internet connection, this can be deactivated or only the app can not be granted access to the internet.

**Transparency** – This procedure step and the function of the server are described in the procedure documentation. The specification, documentation and source code of the server are open, so the server can be checked.

**Unlinkability** – See section 4.7.

## 4.9 Identification of further requirements of the GDPR

A procedure must be implemented for the regular review of the processing activity pursuant to Art. 32(1)(d) GDPR. Specifically this means that a data protection and IT security management must be implemented, with which the controller implements and enforces protection and control measures. In the case of high risks, a data protection officer must be appointed. The task of the data protection officers is to supervise the activities of the executive data protection management, which is integrated into an organisation-wide quality management. The IT and data protection management has the task of identifying deficiencies in good time and remedying them effectively (Rost and Welke 2020). The IT security of all IT components involved must be ensured. To ensure availability, at least the central components must be designed redundantly and backups must be made. We want to underline again to implement redundancy. The authenticity of the servers, clients and services involved must be ensured by using certificates from a public key infrastructure. All communication connections must be end-to-end encrypted so third parties cannot detect whether personal data is being transmitted. The rights of data subjects to information and disclosure, to the consideration of corrections and deletions (cf. Art. 12–22 GDPR) must be effectively implemented by the controller and the used technical components. It must be possible that the data subject herself can trigger the upload of the Health TempIDs in the event of infection.

## 4.10 Designation of the controller

The controler must provide her contact details, as well as those of the representative and the data protection officer.

# Chapter 5

# Lawfulness of processing and accountability

This chapter serves to describe the existing and forthcoming legal grounds for the procedures and processing activities and the allocation of responsibilities for these processing activities.

## 5.1 Lawfulness of processing

In implementing Article 8 of the Charter of Fundamental Rights of the European Union the GDPR provides the conditions for the protection of rights and freedoms of natural persons, which allow for the processing of personal data. This also applies to the processing of personal data in the context of measures taken to contain and control COVID-19, especially when using an app.

Any processing of personal data initially involves interfering with the rights and freedoms of the data subjects and must therefore be justified. The prerequisite for justifying such fundamental rights infringements is compliance with the requirements of the GDPR, the ePrivacy Directive ePrivacy Directive (ePrivacyRL) and the member state regulations based on it; in Germany, these are the BDSG BDSG 2018 and sector-specific law. The individual requirements are set out in the form of the principles of processing in Art. 5 GDPR. These include among others, the lawfulness, necessity and the proportionality of processing as well as the correctness of data and the principle of data minimisation. The controller is responsible for ensuring compliance with the principles relating to the processing of personal data. It is also the controller who, for the purpose of establishing the lawfulness of processing, is obliged to demonstrate that its processing can be based on a legal ground and that it has implemented appropriate measures to protect the data subjects' rights and freedoms.

### 5.1.1 Personal data

The precondition for the material scope of application of the GDPR, and thus for any legal basis for data processing, pertains to the processing of personal data.

Peronal data in Art. 4 no. 1 GDPR means

> "any information relating to an identified or identifiable natural person ('data subject'); an identifiable natural person is one who can be identified, directly or indirectly, in particular by reference to an identifier such as a name, an identification number, location data, an online identifier or to one or more factors specific to the physical, physiological, genetic, mental, economic, cultural or social identity of that natural person;"

Movement and location data (types 1 and 2, see chapter 1.1) that are collected from a terminal device are location data that can be assigned to a natural person by the provider via the telephone number, IP address, MAC address or any other hardware address of the terminal device, or more generally via its network and hardware address.

Contact data are not explicitly defined by the GDPR. Contact data are determined by comparing TempIDs. They are also generated on a person's smartphone and are related to a person as long as they are stored on the smartphone.

Telecommunication data are account and traffic data that can be assigned to natural persons by the provider and are therefore usually personal data.

### 5.1.2 Health data

Over and above the legal preconditions that need to be met to process personal data, the processing of personal data in the categories specified in Art. 9 para. 1 of the GDPR requires (further) special justification. In principle, their processing is prohibited unless one of the exceptions listed in paragraph 2 applies.

Desease data are a special category of personal data according to Art. 9 para. 1 GDPR that indicate something about a person's negative health status.

In the event of a positive test for COVID-19, the TAN sent to the data subject by the competent health authority or institution consitutes health data; the same applies to the health TempIDs of positively diagnosed persons who have received a TAN.

If health data are transferred to the server(s) and if the operator implements an effective isolation procedure (see chapter 8.3), the resulting data are non-identifying infection-indicating data (NIID).

### 5.1.3 Processing

The notion of processing[1] is illustrated in Art. 4 no. 2 GDPR using examples. According to these examples, processing means

> "any operation or set of operations which is performed on personal data or on sets of personal data, whether or not by automated means, such as collection, recording, organisation, structuring, storage, adaptation or alteration, retrieval, consultation, use, disclosure by transmission, dissemination or otherwise making available, alignment or combination, restriction, erasure or destruction;"

Art. 4 no. 2 GDPR lists various processing operations that reflect the life cycle of personal data. This indicates that any handling or use of personal data should constitute processing as understood by the GDPR.

When processing personal data in the context of the use of CA, a distinction can be made between various processing activities and processing operations (see chapter 4) The distinction between processing activities and processing operations is a prerequisite for correctly determining responsibility: if the processing operations are distinct operations or sets of operations, they may require different legal justifications and may give rise to separate or joint responsibilities. The following distinctions can be made for the purposes of legal classification:

- Using the app on a smartphone (end device)
    ∘ Processing on the sender's end device
        ∗ Generation and transmission of TempIDs to other app users

---

[1] See also chapter 4.

- - - * Collection of TempID tokens received from other app users
      * Storage of TempIDs received from other app users
    ○ Transmission of the health TempIDs to the server(s)
  – "Entering" a TAN for authentication = status change in the app to "infected" = status change of TempIDs to health TempIDs in the app
  – Transmission of health TempIDs to the server
  – Server
    ○ Anonymisation of the health TempIDs into non-identifying infection-indicating data (NIID)
    ○ Provision of non-identifying infection-indicating data (NIID) for retrieval
    ○ Storage of telecommunications data
  – Processing on receiver device
    ○ Retrieving non-identifying infection-indicating data (NIID) from the server
    ○ Matching of the TempIDs on your app against the downloaded NIID = Creating the contact data
    ○ Calculation of the risk score based on the contact data
    ○ Notification: Infection warning, contact with infected person yes/no

The question of whether the operator of the app or the operator of the server can access the encrypted or pseudonymised data is irrelevant in terms of whether personal data are processed (see EDPB and EDPS 2019, Para 8). In order for the processing of personal data to be confirmed, it is sufficient for the TempIDs to be generated on the user's terminal equipment. The fact that the tokens are sent in encrypted form via a secure network does not change the personal nature of the tokens. Even with encrypted personal data, the personal reference remains intact.

The necessity to process personal data can only be determined in relation to the purposes for which they are intended (see chapter 5.4.3). The purposes are defined by the controller.

## 5.2 Accountability

The processing of personal data can only be lawful if a body responsible for the processing can be identified. Encroachments on the rights and freedoms of natural persons due to the processing of personal data should not take place in a vacuum (see Recital 78 GDPR). In Art. 5 para. 2 in conjunction with Art. 4 no. 7 GDPR, the GDPR assigns responsibility to the body, which alone or jointly with others determines the purposes and means of processing. The entity in question – the natural or legal person, authority, institution or other body thus identified – is responsible under Art. 5 para. 2 GDPR for ensuring that the principles of processing under Art. 5 para. 1 GDPR are complied with and must be able to demonstrate this. According to Art. 24 para. 1 GDPR, the controller must verifiably take technical and organisational measures (Recital 74 GDPR) for compliance with the principles of Art. 5 para. 1 GDPR that ensure adequate protection of the rights and freedoms of the data subjects, for the entire processing operation. Pursuant to Art. 35 GDPR, the controller is also obligated to implement a DPIA.

The attribution of controllership is a consideration of the actual, functional role that the controller plays in the processing (Article 29 Data Protection Working Party 2010, p. 9). A formal assignment of controllership can be made by legal assignment in exceptional cases.

According to Art. 4 no. 7 GDPR, a controller is defined as

> the natural or legal person, public authority, agency or other body which, alone or jointly with others, determines the purposes and means of the processing of personal data; where the purposes and means of such processing are determined by Union or Member State law, the controller or the specific criteria for its nomination may be provided for by Union or Member State law;

The controller may be any natural or legal person, public authority, agency or other body. Thus, all actors involved in the processing (see chapter 2.1) can be considered controllers. Since there are no legal provisions on CA or their use, it is important to determine whether and to what extent each body decides on the purposes and means of processing.

The purpose of processing is any result that guides the processing. Pursuant to Art. 5 para. 1 lit. b GDPR, purposes must be established prior to processing and must be explicit and legitimate. The means are defined as how the result is to be achieved.

In this regard, the actual power to decide with respect to the essential elements of the processing is the decisive criterion. The controller must always decide both on purposes and on the essential means. A capacity to decide on the purposes alone is not sufficient. Decisions on essential means are decisions on which data are processed, on the time period of the processing and on who may have access to the data. For instance, decisions on the selection of specific hardware and software are not necessarily essential. Decision-making power may also arise on statutory grounds if a law lays down criteria for determining what constitutes actual decision-making power. If the purpose of the processing emerges due to a statutory assignment of tasks, this may also be an indication of actual decision-making power.

For the de facto decision-making power, the actual circumstances of the processing operation must be considered. This can be done by asking "Why is this processing being carried out?" and "Who initiated it?" (Article 29 Data Protection Working Party 2010). It should be noted that a processing operation may consist of different processing operations, either in isolation or when linked in such a way that they can only be considered a coherent processing operation when combined with each other. In this respect, a distinction must be made between the different processing operations. An indication of a single processing operation is given by the purpose of the processing. Processing operations serving the same purpose should normally be regarded as one operation or a set of operations. The determining factor is the decisive influence on the specific processing context. The capacity to decide on the grounds for processing and thus on its purpose is a characteristic of actual decision-making power in respect of processing.

Being natural persons, users are not excluded from the concept of controllership from the outset. With regard to the purposes of processing and the provision of the CA to the users, the possibilities for users to exercise decision-making power with regard to the "how" of processing and the technical design take a back seat, or are in fact non-existent due to a lack of configuration options. Although users can decide on the "if" of the processing by installing the CA, they cannot influence the purpose of the processing. The fact that processing steps such as the generation of TempIDs or matching takes place on the user's terminal equipment is not in itself decisive for the allocation of decision-making power over the purposes and means (cf. the discussion on DRM systems that run on the user's terminal equipment and enforce the rights of the rights holders there (Becker et al. 2003)). The exchange of TempIDs between users does not constitute a processing operation in itself, as the purpose in question is a purpose other than mere exchange, namely information matching. Therefore, the users are not independently responsible for this.

If, for statutory reasons, the CA is operated by a public authority, e.g. the Federal Ministry of Health (BMG), or by another public body, e.g. the Robert Koch Institute (RKI), the relationship with the users will be characterised not only by an informational asymmetry of power, but also by a relationship of subordination that indicates that the public body is responsible. Where the operator decides why and how the data are processed, the de facto power of decision making by the operation lies with the operator, who is thus responsible for the processing. This also applies if a private body is acting as operator of the app. It is irrelevant whether this is also the app manufacturer. The app manufacturer only has decision-making power if it decides on the essential means and purposes of processing. Although the app manufacturer configures the technical system and thus has an influence on how it works, it does not ultimately decide on the purposes of the actual processing for which the app is used.

Where the CA and the server(s) necessary for the use of the service are maintained by different legal entities, controllership depends on who has the actual power to decide regarding processing operations. This depends on the type of CA (see chapter 1.1). If the matching takes place on the server(s) of an operator and the operator decides on the procedure for matching and/or notifying the recipient (types 1 and 2), or if the operator stores the ID lists and further processes them, for example to calculate further results that may be aggregated (e.g. statistics), it has actual decision-making power for these processing operations. If matching is done on the terminal equipment (type 3), this indicates actual decision-making power and thus responsibility on the part of the CA operator. If the server is operated by another legal entity, processing on behalf of the controller by the operator of the server can be considered in this constellation.

It is not necessary for the controller to have access to the personal data. All that is required is that the controller can determine that personal data are processed on the user's devices in connection with the app as well as deciding on which personal data are processed in this respect.

Furthermore, the controllers' control over the processing does not have to be exercised over the whole procedure; it is sufficient that it relates to specific processing operations and that other operations are the responsibility of another body (see ECJ, Fashion ID C-40/17, ECLI:EC:C:2019:629, paragraph 74).

The operators of the CA and the server(s) are regarded as jointly responsible when they jointly decide on the purposes and means of the processing (Hartung Kühling and Buchner 2018, Art. 26 Rn. 11).

The users cannot be held responsible within the context of joint control with the operators. However, the exclusion of the material scope of application of the GDPR, which could be considered if the processing of personal data by natural persons was carried out for the exercise of purely personal or household activities in accordance with Art. 2 para. 2 lit. c GDPR, cannot be regarded as a justification. The exchange of TempIDs is not a purely personal or family activity, but serves the purposes of infection control or pandemic containment (see chapter 4.2). Ultimately, however, this question may remain unresolved, because, on the one hand, the affirmation of exclusively personal or family activities does not rule out that the operator may have responsibilities (cf. Recital 18 GDPR) and, on the other hand, users have neither a legal nor a factual influence on the purpose and the decision on how personal data are processed on their devices and are therefore not to be regarded as controllers.

## 5.3 Lawfulness of processing

The principle of lawfulness presupposes, according to Art. 6 (1) GDPR, the existence of legal grounds for the processing. A legal basis must be determined for each processing purpose.

The legal bases of the GDPR are conclusively regulated in Art. 6 para. 1 sentence 1 GDPR. The following provisions may serve as the legal bases for the processing activities described in Chapter 5:

– Consent under Art. 6 para. 1 sentence 1 lit. a GDPR;

– Contract under Art. 6 para. 1 sentence 1 lit. b GDPR;

– Legal obligation under Art. 6 para. 1 sentence 1 lit. c GDPR;

– Legal task according to Art. 6 para. 1 sentence 1 lit. e GDPR in conjunction with the Federal Infection Protection Act;

– Vital interests of the data subject or another person concerned pursuant to Art. 6 para. 1 sentence 1 lit. d GDPR;

– Legitimate interest under Art. 6 para. 1 lit. f GDPR.

The choice of the legal basis depends on the role of the controller, the purpose of processing, and the context of the processing operation (European Data Protection Board 2019a, Rn 18). If the responsibility for the processing activity lies with a public body such as the RKI or the BMG, Art. 6 para. 1 sentence 2 GDPR does not provide legitimate interest as a legal basis. Furthermore, in order to determine the appropriate legal basis, the processing purpose and the context of the processing must be taken into account.

For the processing of personal data, including the processing of health data by health authorities or other bodies acting on their behalf, Art. 6 para. 1 sentence 1 lit. c and e in conjunction with Art. 6 para. 3 lit. a, Art. 9 para. 2 lit. g and lit. i GDPR come into consideration in particular. The prerequisites for this include that the processing of personal data falls within the statutory remit of the authority in accordance with Art. 6 (3) sentences 2, 3 GDPR, that the processing of personal data is necessary (see chapter 5.4.3) and that the further provisions of the GDPR and sector-specific law are complied with.

### 5.3.1 Consent, Art. 6 Abs. 1 S. 1 lit. a GDPR

The information on the CA provided in public announcements has so far included a reference to the fact that citizens can decide to use it voluntarily. To date, there has been no legal regulation of the use and application of a CA that would allow voluntary use. Any processing of personal data in connection with the use of the CA in conformity with the law on the basis of the granting of data protection consent can only take place on condition that it complies with the legal provisions. There are considerable concerns about this in terms of both the normative requirements and the operational implementation.

With regard to the possibility of invoking consent Art. 6 para. 1 sentence 1 lit. a GDPR as the legal basis for processing activities , the following aspects must be taken into account:

- The use of the CA is *one* part of the procedure (see chapter 4 within the processing activities and purposes).

- Consent to the use of the app and to the processing of the consenting persons' personal data does not automatically imply consent to the (entire) procedure.

- Consent to use the app must be separated from consent to the transmission of the TAN, i.e. authentication after a positive test, to the server and transmission of an individual's own health TempIDs.

- Consent management is essential.

If the processing of personal data is to be based on consent, the requirements of Art. 6 para. 1 sentence 1 lit. a in conjunction with Art. 7 and Art. 4 No. 11 GDPR must be met. According to Art. 4 no. 11 GDPR, the consent of the data subject means

"any freely given, specific, informed and unambiguous indication of the data subject's wishes by which he or she, by a statement or by a clear affirmative action, signifies agreement to the processing of personal data relating to him or her;"

Voluntariness presupposes a genuine choice for the data subject Article 29 Data Protection Working Party 2018, p. 5; only then can it fulfil its protective function. In order to assess the options, the context in which consent is to be given must be considered in more detail. The purpose of the CA is to enable notification of an increased risk of infection due to contact with a person who has tested positive, so that this person can be quarantined and self-tested at an early stage to contain the pandemic. The argument against voluntariness is that there is a clear difference in power between the controller and the data subject. This is classically the case in the relationship between citizens and public authorities, which is both superior and subordinate (Recital 43 GDPR). If consent is given to a public authority, it is generally assumed that it is not given voluntarily (burden of proof to the detriment of voluntariness in Recital 43 p. 1 GDPR). Rather, it must be demonstrated in each individual case that the imbalance that typically exists in the given situation does not apply. Special requirements must therefore be made when granting consent to an authority (buck in Specht and Mantz 2019, 570 Rn 27 sq.). This imbalance is to be counteracted by the fact that the use of the app is not mandatory and does not entail any standardised, direct legal consequences.

However, the absence of directly normative disadvantages in the case of non-use of the app does not immediately mean that consent to the data processing associated with the use of the app is voluntary within the meaning of Art. 7 para. 4 GDPR. In the context of the processing, it should be noted that, among other things, the question of loosening lockdowns by government representatives has been explicitly or at least effectively linked to the use of the app and the widest possible use by the population (at least 60 %) (Neuerer 2020). In fact, the consequence of non-use would be the prolongation or even further restriction of public life and the restriction of fundamental rights (freedom of movement). The voluntary nature of the consent is thus counterbalanced by an expected burden. The particular feature in this case is that the expected actions by public authorities would not be directed against the individual citizen but would take the form of general restrictions of freedom applying to all citizens without distinction. Against this background, the possibility that social pressure may arise to a not inconsiderable extent to behave in accordance with the expectations of the social environment, the state or the app operators and to use the app accordingly must be taken into account, because even legally non-sanctioned behaviour can call the voluntariness of the behaviour of the data subject into question if it triggers an abstract internal compulsion (Heckmann/Paschke in Ehmann and Selmayr 2018, Art. 7 margin no. 51). The desire to return to a more or less normalised public life is understandable. The citizens will hardly be able to close their minds to this.

In order to promote the purpose of the data processing under consideration here, an individual relaxation of the restrictions on freedom could then be combined with the use of the app, for example by making access to playgrounds, the workplace or school attendance dependent on the use of the app. This social pressure also seems to be desired by the authorities (Randow 2020).

In addition, there is no realistic alternative to using the app and consequently no possibility to refuse the processing that goes with it. A voluntary testing of large parts of the population is currently not possible due to capacity problems. An infected person is already contagious before COVID-19 symptoms appear. If the contact persons for restricting the pandemic are to be found in time, this is traditionally done by the health authorities by means of a questionnaire and telephone. A digital system that identifies and informs the contacts would be comparatively faster. This further increases the social pressure on individuals to participate in the process. Further legal or at least de facto disadvantages could be that access to tests for symptoms may be made dependent on proof of contact with a confirmed infected person and that the app would be the only available means of establishing credibility. It is not just these direct consequences of non-use that would have an impact on deciding whether or not to use the app; inauspicious developments in this regard are already pointing in this direction. From a purely legal point of view, the GDPR addresses these concerns by categorically denying the voluntary nature of consent vis-à-vis the authorities and Recital 43 p. 1 GDPR clarifies that in special cases, where there is a clear imbalance between the data subject and the controller, particularly where the controller is an authority, and it is therefore unlikely, in view of all the circumstances in the specific case, that consent was given voluntarily, consent cannot provide a valid legal basis. Seen as a whole, the app is a component in the context of state pandemic control and thus an administrative act that aims at proportionate intervention to avert further restrictions on freedom. In this respect, the necessity of using the app can be seen in the context of an intervention by administration. In this context, consent is also widely rejected as a legal basis in the literature on data protection (Heckmann/Paschke in Ehmann and Selmayr 2018, Art. 7 margin no. 53, see also Hoffmann 2017 for a critical view of the area of social law).

Furthermore, a prerequisite for an effective consent is its specificity. According to Art. 5 para. 1 lit. b GDPR, personal data may only be collected for specified, clear and legitimate purposes. The fight against the pandemic undoubtedly constitutes a legitimate interest, but in its generality the purpose remains undefined. The determination of purpose must in principle be as precise as possible and it must be ensured that personal data are not processed for purposes that the data subject did not expect when they were collected (see Kühling and Martini 2016, p. 51: "narrow understanding").

If the purpose of providing the CA is to provide rapid information to persons who have come into contact with infected persons, no further data processing may be carried out beyond this purpose (see chapter 4.2).

The purpose for which consent is to be given must be specifically described. This is to ensure that the controller does not subsequently add other purposes or extend the processing operations. Therefore, the information to be provided to the data subject about the purpose of the consent must be clearly separated from further information concerning the processing (Article 29 Data Protection Working Party 2018, p. 11).

Consent should be given for interrelated processing operations. If different purposes are combined, consent must be obtained for each of the purposes (Recital 32 GDPR). A distinction must be made between two processing activities for the use of the app. Firstly, the exchange of TempIDs between the apps and secondly, the transmission of TAN and health TempIDs to the server in the event of a verified infection (cf. chapter 4). If the user does not pass on the information to the server, the purpose and form of the declaration of consent will be decisive. If the user does not forward the TAN, it is to be assumed that he/she does not agree to the further processing of the information, which concerns a positive test event. Both activities must be triggered by an action of the user. If one understands the exchange of TempIDs as a prerequisite for being informed about contact with an infected person, consent can refer to both actions. To the extent that the user is to be given the possibility to freely decide whether to pass on the information on their positive test result, this choice should also be clearly expressed in the informed consent form.

For consent to be effective, the controller must prove that it is possible to object to the consent and that the consent can be withdrawn without negative consequences. Although withdrawal only has consequences ex nunc, and thus from the time of its declaration, it must also be possible to withdraw from the procedure from that time onwards. With regard to the use of the CA, it has not yet been explained what consequences withdrawal of consent would have for the data subject. In particular, there is a lack of information on how to proceed with withdrawal in case of infection. It has also not been clarified whether withdrawal is technically possible at all using the app or how it is implemented in the procedure, because withdrawal of non-identifying-infection-indicating data (NIID) on the server also leads to deletion of this data on the server, which runs the risk of undermining the effect of the intended separation, i.e. anonymisation, which protects fundamental rights. Against the background of conflicting requirements, a legal assessment can only be made with regard to a definite operative implementation.

Consent must be given in an informed manner. To this end, the controller must provide the data subject with the following information: the identity of the controller, the purposes of the processing for which consent is to be obtained, the categories of data to be processed, the right of withdrawal if the data are used for automated decision making under Art. 22 para. 2 lit. c GDPR, and the use and risks of transfers to third countries that do not have an adequate level of data protection. This information should, to the extent possible, be provided in clear and simple language and be easy to find and not hidden between other text (Article 29 Data Protection Working Party 2018, p. 14). However, informed consent can only be given if the data subject is also able to assess the scope of the consent, i.e. the consequences including the risks of the processing (Buchner/Kühling in Kühling and Buchner 2018, Art. 4 No. 11 recital

5-12).

If, within the framework of the legal basis of Art. 6 para. 1 sentence 1 lit. f GDPR, the controller must weigh their legitimate interests against those of the data subject and his/her fundamental rights and fundamental freedoms, then in the case of an equivalent level of protection conveyed by the legal bases, the information provided by the controller must, within the context of consent, put the data subject in a position to weigh up these options. In order not to shift the responsibility for weighing up options and thus for assessing the risk completely onto the data subject, the fundamental rights risks for the data subject should be presented in clear and simple language, starting from the description of the procedure, in order to enable the data subject to weigh the fundamental rights risks against the objective pursued (Rost 2018b). Such a presentation in clear and simple language hardly seems possible in view of the complexity of the procedure and the effects of using the CA, as described in chapter 4 and 7. The informational power asymmetry between the controller and the data subject caused by the processing is thus not only not balanced, but rather reinforced. In the end, this imbalance could only be addressed by a statutory regulation by which the risks of the processing would be balanced out.

Consent must be given in a clear manner. In the sense of data protection by default settings (Art. 25 para. 2 GDPR), a factory pre-installation of the app may only take place if it must be activated by the user. In order to ensure the possibility of intervention by withdrawal, the app should be easy to deactivate or uninstall.

The granting of consent by children requires special measures. It must be ensured that children under the age of 16 can only use the app with the consent of a parent or guardian. The authorization must be proven.

Consent pursuant to Art. 5 para. 3 ePrivacyRL is also not required because only data from the terminal device that is necessary for the provision of the service is accessed.

### 5.3.2 Contract, Art. 6 para. 1 S. 1 lit. b GDPR

In principle, a contractual arrangement for usage between a private app operator and a user of the app pursuant to Art. 6 para. 1 sentence 1 lit. b GDPR can also be considered in order to transfer data to third parties or other users in connection with consent pursuant to Art. 9 para. 2 lit. a GDPR for the processing and transfer of health data in the case of server push data.

Theoretically, all users could conceivably commit themselves mutually to carry out the respective necessary data processing and to accept the processing of data by third parties. Apart from unclear liability risks, the main reason for not opting for a contractual arrangement would be that, in the case of the operators being controlled by a public body, a public-law contract in the sense of Sect. 54 VwVfG would have to be concluded, which, according to § 57 VwVfG, requires the written form. This formal requirement is likely to be completely impracticable in the context discussed here.

### 5.3.3 General requirements of lawfulness

Since the use of the app should not be based on the consent of the user due to the lack of voluntariness, since both the liability risks and formal requirements speak against a contractual solution, and since there are many indications that the CA will fall under the Medical Devices Ordinance (European Commission 2020a), legislators are well advised to consider a legal regulation for the procedure and use of the CA.

Even if there is a statutory provision, it cannot be ruled out that the use of the CA itself – generally or only for certain groups, such as non-risk groups – will be on a voluntary basis.

A legal regulation may also provide that, for example, members of risk groups who typically do not have a smartphone are equipped with a suitable device on which only the CA has been installed and is running.

In particular, a legal regulation of data processing allows the implementation of a legislative DPIA according to Art. 35 para. 10 GDPR already at the time of regulation. Legislators should make use of this possibility.

### 5.3.4 Fulfilment of a legal obligation, Article 6 para. 1 lit. c GDPR

Art. 6 para. 1 sentence 1 lit. c GDPR covers the legal obligations resulting from legal provisions. The obligation stipulated in the legal provision must relate directly to data processing (Recital 45 GDPR). The duty must be directly addressed to the controller under data protection law, i.e. to the body that determines the purposes and means of processing. A concrete legal obligation to operate a CA does not currently exist. Obligations to recognise transmissible diseases and infections in persons and to combat and contain them are regulated in the Infection Protection Act (IfSG). SARS-CoV-2 is a pathogen, i.e. an agent capable of reproduction, a virus, which can cause an infection and transmissible disease, COVID-19, in persons.

According to Sect. 3 p. 1 IfSG, informing and educating the general public about the dangers of contagious diseases and the possibilities for their prevention are public tasks that are regulated in the Infection Protection Act. The IfSG aims to prevent diseases that, due to their high risk of infection, can spread rapidly to large population groups and, in combination with their severity, thus represent a major risk for many people.

Section 4 para. 1 sentence 1 IfSG appoints the Robert Koch Institute (RKI) as the national authority for the prevention of communicable diseases and for early detection and prevention of the further spread of infections. According to Sect. 4 para. 1 p. 2 IfSG, the tasks of the RKI also include the development and implementation of epidemiological and laboratory-based analyses as well as research on the cause, diagnosis and prevention of communicable diseases. Further tasks are stipulated in Sect. 4 para. 2 IfSG; according to it, the RKI, in consultation with the respective competent federal authorities, should prepare guidelines, recommendations, leaflets and other information as a measure of preventive health protection to prevent, detect and prevent the further spread of communicable diseases (Sect. 4 para. 2 No. 1 IfSG) and should epidemiologically analyse the data on notifiable diseases and notifiable evidence of pathogens that have been transmitted to it under this act and under Section 11 para. 5, Section 16 para. 4 of the IGV Implementing Act (Section 4 para. 2 No. 2 IfSG). These are general descriptions of tasks. However, Art. 6 para. 1 sentence 1 lit. c GDPR presupposes that the purpose of processing is directly derived from the legal obligation (Recital 45 GDPR). A duty to provide app-based information to persons who have come into contact with a person who has tested positive does not, however, emerge from the current text of Sect. 4 IfSG.

Section 4 (1a) IfSG deals specifically with the epidemic caused by SARS-CoV-2, but this provision only provides for a reporting obligation and also does not regulate the use or authorisation to operate a CA.

From Sect. 6 in conjunction with Sect. 9 IfSG only a reporting obligation for infectious diseases results, but no concrete obligation to warn possibly infected persons or to operate a CA.

Section 13 IfSG authorises the RKI to monitor communicable diseases in cooperation with selected institutions of health care or of preventive health care through so-called sentinel surveys. These serve the collection of anonymous, random data, which are intended to draw conclusions on the spread of infectious diseases. The purpose of operating a CA, to enable a person to be warned as early as possible, is not covered.

Section 14 IfSG regulates an electronic reporting and information system. The purpose of this system is to enable the electronic registration of facts subject to the obligation to report and notify under the IfSG by persons and bodies subject to the obligation to report. This standard does not therefore cover the purposes of processing personal data by a CA, because the legal obligation must make it necessary for the CA to process the personal data in order to fulfil the legal obligation. This is intended to ensure that the responsible party does not take the legally prescribed objective as an excuse to process further personal data or to process them for other purposes (Article 29 Data Protection Working Party 2018, p. 11).

However, the clear and precise definition of at least the purpose of processing is a prerequisite for creating a legal obligation to process personal data (EC 41 GDPR), even if it is not absolutely necessary for each individual processing operation to be regulated by a specific law. Recital 41 GDPR clarifies in this respect that the constitutional order of a Member State concerned remains unaffected. In the legal tradition of the German Basic Law, this also includes the principle of essentiality and specificity, which is why German legislators will in any case be required to regulate the essential processing objectives and steps of the CA itself. None of the standards of the IfSG for a CA procedure or the operation of a CA fulfil these requirements.

According to Art. 6 para. 3 sentence 4 GDPR, the legal basis must pursue an objective in the public interest and be proportionate to the legitimate aim pursued. Examples of rules result from 9 para. 2 lit. h and i and 23 lit. e GDPR. The provision must be determined in the sense of Art. 5 para. 1 lit. b GDPR, i.e. its content must be clearly and precisely derived from the legal provision.

Section 13 para. 1 sentence 1 authorises the Federal Government and the Länder to carry out further forms of epidemiological surveillance of communicable diseases and to regulate them by ordinance.

If the RKI is to be entrusted with the operation of the CA itself or with the possibility of involving a contract processor, the legal prerequisites for this would have to be created in the IfSG (Section 14 (8) IfSG) and further provisions, in particular technical specifications, should be regulated in an implementing regulation. In particular, it must be determined "which functional and technical specifications, including a safety concept, must form the basis of the electronic reporting and information system" (§ 14, Subsection 8, No. 4 IfSG), which satisfies the requirements of the GDPR. The responsible ministry has not yet issued such an ordinance. However, it is necessary.

### 5.3.5 Performance of a task in the public interest, Art. 6 para. 1 sentence 1 letter e GDPR

If the operation of a CA is to be based on Article 6(1)(1)(e) of the GDPR, the purposes of operating a CA must be in the public interest or in the exercise of official authority delegated to the controller.

The purpose of containing the COVID-19 pandemic by providing information as early as possible about contact with a person who has tested positive and thus early warning of possible infection is undoubtedly in the public interest.

Control can be exercised in public authority by a public administration, by another public body or by a body governed by private law. This requires the transfer of an appropriate power to enforce the task existing in the public interest.

The further requirements of Art. 6 para. 1 sentence 1 lit. e GDPR are identical to those of the legal basis in Art. 6 para. 1 sentence 1 lit. c GDPR. Thus, the task in the public interest must be defined by law. Although the warning of infection with an infectious disease and containment of a pandemic is in principle covered by the Infection Protection Act, the operation and use of an app to provide information about contacts with an infected person is not regulated by law.

A legal regime which allows the processing of personal health data in the framework of a CA type 3 (see chapter 1.1) with the purpose of limiting the spread of COVID-19 in order not to overburden the health care system must contain proportionate rules for this purpose, which minimise the processing of personal data, and in particular the processing of health data, and keep it at an appropriate level. Such processing must not lead to the general and systematic collection, storage and further processing of personal data. The method of processing must be suitable for achieving the purposes and there must not be equally suitable, less intrusive means available for achieving the purposes. The specific role of the authority as controller of the data processing must not be extraneous and should fit within its remit.

### 5.3.6 Protection of vital interests, Art. 6 para. 1 sentence 1 lit. d GDPR

Art. 6 para. 1 sentence 1 lit. d GDPR could also be considered as a legal basis that is intended to protect the vital interests of the persons concerned or another natural person. This is a catch-all provision as a substitute for consent if consent cannot be granted under the specific circumstances, but its granting could be expected (cf. Albers in Wolf and Brink 2017, Rn. 37), comparable, for instance, to the anticipated consent in the case of unconsciousness in the field of medical treatment. However, such a constellation is not given in the present case. Nor does the data processing serve to protect the vital interest of another natural person, because the app and its use do not protect against infection. It is also not necessary, because, as Recital 46 GDPR states, this element of permission should only apply if no other legal basis is obviously relevant, and such a legal basis is present here with the performance of a task in the public interest pursuant to Art. 6 para. 1 sentence 1 lit. e GDPR.

## 5.4 Proportionality

All legal bases require that the processing of personal data must be proportionate or adequate in relation to the purposes for which they are processed. The legal bases of the GDPR formulate this in part explicitly, for example in the area of creating state processing regulations, cf. Art. 6 para. 3 sentence 4 GDPR, and in part this simply follows from the general provisions of Art. 5 para. 1 GDPR.

A processing operation is only proportionate if without it the purpose cannot be achieved, or cannot be achieved reliably or only with disproportionate effort. This is to be assessed according to objective criteria. There must be a connection between the data and the purpose pursued by the processing (Heberlein in Ehmann and Selmayr 2018, Art. 6 marginal no. 23).

The assessment of whether or not this connection is present in the individual case is demanding. It presupposes, first, a description of the semantic content of the infor-mation and of any processing which may be envisaged with it, second, a description of the purpose of the processing and, third, a description of the degree of dependence with which the attainment of that purpose depends on the actual processing of the personal data concerned, in a form which makes that degree of dependence objectively ascertainable and rationally debatable (Podlech 1982, 455f.).

In cases where different conflicting fundamental rights are balanced against each other, this balancing must always be done in relation to very specific processing operations and their concrete forms. Before a political decision can be taken on the statutory regulation by which the concrete purpose is set and legitimised and the processing operations to be used for it can be determined, the purposes, the processing operations and the connections between the two must be discussed, not only on a legal but also on a technical basis, because both the determination of the required specificity of the purpose and the processing operations represent a challenge (see Härting 2020).

The principle of proportionality requires that the processing of personal data

- serves a legitimate purpose,
- is suitable to achieve this purpose,
- is necessary to achieve this purpose (i.e. there is no milder, equally effective means), and
- is appropriate, i.e. proportionate in the narrower sense.

A processing operation is legitimate if its purpose is basically within the scope of the tasks entrusted to the state. It is suitable if it can in principle serve this purpose causally. It is necessary if no milder means is suitable to serve this purpose comparably effectively. Processing is appropriate – or proportionate in the narrower sense – if the seriousness of the encroachment on fundamental rights is not disproportionate to the weight of the reasons justifying it in an overall assessment. This requires a balancing of legal interests. This can never be dealt with in a purely legal manner, but always has a political dimension.

### 5.4.1 Legitimate purpose

The overall purpose (see chapter 4.2) of corona apps is usually indirectly related to the protection of life and the physical integrity of persons during a pandemic; thus, seen as a whole, it serves to contain and control the pandemic, specifically to delay new infections in order not to burden the health care system beyond its capacity. Individuals cannot pursue this goal on their own, so it is a legitimate purpose that can serve as a basis for a legal obligation or the performance as a task in the public interest (BVerfG 2020, rn 14).

Since the balancing has to take place in the specific case, the following purposes are to be considered as examples in the context of this DPIA

- Purpose A is to inform potentially infected (exposed) persons, i.e. to warn people who have had contact with infected persons so that they can be quarantined.
- Purpose B is to generally review compliance with curfews in order to evaluate political action.

The purposes given as an example here can be pursued by using different technologies or without the use of technology. However, the purpose in itself is not a technical question.

### 5.4.2 Suitability

The question of the suitability of processing has a technical dimension, because if a certain technology cannot serve the purpose at all, it must not be used.

Purpose A: A technical evaluation of GPS or mobile phone metadata shows that these data are not accurate enough to identify epidemiologically relevant contact events. These technologies are therefore excluded. Short-range technologies such as Bluetooth, on the other hand, are suitable because, among other things, they are even intended for distance measurements in the metre range.

Purpose B: GPS or mobile phone metadata would be technically suitable for the generation of the of general population movement statistics required in example B. Short-range technologies, on the other hand, are only suitable to a limited extent because they do not easily provide a location reference. Aggregated data – i.e. summarized and purely statistical data that can be calculated from GPS, mobile phone metadata or close-range data – would also be suitable.

### 5.4.3 Necessity

In many of the legal bases of Art. 6 (1) sentence 1 GDPR under data protection law, necessity is one of the essential prerequisites. When considering the necessity of processing, a technical dimension must be taken into account. If an end can also be achieved by "milder" technical means because it has a lower intensity of interference with fundamental rights for technical reasons, the milder means must be chosen and the means currently under consideration may not be used. In order to evaluate whether there is a milder means or what a milder means can be, a technical consideration will usually be necessary. In this consideration, the requirements of Article 25 GDPR (data protection by design) must also be taken into account. This means, among other things, that state-of-the-art technologies must be taken into account and the principle of data minimisation must be observed.

Purpose A: The use of short-range technologies such as Bluetooth could be necessary if, for example, the task of health authorities to detect infection chains quickly and efficiently could be fulfilled much more quickly and accurately by using an app than through questionnaires and telephone calls.

Purpose B: No individual personal data are required for the general verification of compliance with curfews, and therefore only aggregated statistical data can be regarded as a milder means. More detailed data, such as specific contact event data, individual GPS measurements or other location data are not applicable here, as they are more intrusive but not more effective.

### 5.4.4 Appropriateness

Processing is only adequate for achieving a purpose if the specific balance of interests is in favour of the controller in the context of a purpose and means ratio. The interests of the data subjects must therefore be weighed against the interests of the controllers in the processing. This weighing up must also take into account the effects and "side effects" on society as a whole. These can be medical issues, but also social, economic or psychological ones, whereby these are also linked to each other.

Purpose A: From a technical point of view, purpose A – if at all – can be achieved with short-range technologies. In this example, appropriateness depends on technical implementation details. Even small changes can lead to a different outcome.

Purpose B: From a technical point of view, purpose B can be achieved with aggregated data alone. In any case, it is not possible to make general statements beyond that, because the intensity of interference depends very much on the concrete technical implementation, as the current discussion about the PEPP-PT framework and the decentralised DP-3T implementation shows.

Last but not least, the issue of whether the concrete effect of the use of the app is at all proportional to the restricted rights is relevant. This is a particularly delicate issue in the case of experimental apps such as the proposed corona tracing apps, as their benefits cannot be estimated at present. The – so far only theoretically modelled – effect can only be expected when used by at least 60% of the population. Findings from Singapore may be instructive in this respect. However, only 13% of the people there had installed the individualized TraceTogether-App. Additional aspects relevant for consideration are discussed in detail in the threshold value analysis (see chapter 6).

## 5.5 Right to information

In order to comply with the principles of processing under Art. 5 (1) GDPR, the controller must ensure that the processing is carried out in a way that is comprehensible to the data subject. Transparency can be ensured by the fulfilment of the information duties. The controller must therefore provide data subjects with all information pursuant to Articles 13 and 14 and all notifications pursuant to Articles 15 to 22 and Article 34, which relate to the processing in a precise, transparent, comprehensible and easily accessible form in clear and simple language (Art. 12 para. 1 sentence 1 GDPR). This should be done in a data protection statement, which should be accessible from the app.

More specifically, this includes information about the purposes of the processing, the measures used for those purposes, in particular the duration of the storage of personal data, data transmissions and their recipients, and how data subjects can effectively exercise their rights vis-à-vis the controller, including through recourse to the competent data protection supervisory authority.

## 5.6 Technical and organisational measures

Art. 24 para. 1 GDPR requires that the controller – taking into account the nature, scope, circumstances and purposes of the processing and the varying degrees of likelihood and severity of the risks to the rights and freedoms of natural persons – implements appropriate technical and organisational measures (TOMs) to ensure and provide proof that the processing is in compliance with data protection law. It is data-protection compliant if it fulfils the data protection principles in accordance with Art. 5 GDPR, effectively implements the requirements in accordance with Art. 32 GDPR, effectively guarantees the rights of data subjects in accordance with Art. 12 to 22 GDPR, implements this through technical design and default settings and is accompanied by data protection management.

For more on the technical and organisational measures to be implemented, see chapter 8.

# Chapter 6

# Performing the threshold value analysis

The purpose of a threshold analysis is to determine the level of risk of a processing activity for data subjects. If there is a high risk to the rights and freedoms of data subjects, Art. 35 GDPR provides that the controller must carry out a data protection impact assessment.

The GDPR distinguishes between two risk levels, namely a low/normal risk and a high risk. Correlating to the level of risk, the processing must a) be designed with particular care with regarding the intensity of the fundamental rights intervention, and b) increased effective protective measures must be carefully specified and implemented, as well as operated in a verifiable and controllable manner. Since the mere fact of processing personal data normally causes a normal risk for the data subjects, protective measures must be taken in any case for used means of data processing. These measures must ensure that the operational requirements for a processing activity according to the Articles 5, 24, 25, 32 GDPR are met.

Three sets of criteria are used to determine the risk level for the processing activity in question: a) Art. 9 GDPR, b) Art. 25 GDPR and c) the list from Working Paper 248 of the Art. 29 Group (Article 29 Data Protection Working Party 2017)

a) Art. 9 GDPR lists special "categories of personal data", which includes "health data". The processing of this type of special data is generally prohibited and only permitted under the conditions laid down in the Act. For practical implementation, this means that processing this type of data carries a high risk that the increased requirements of the DPA will not be met more intensively in the interests of the data subjects.

The corona app (CA) itself collects only distance data between smartphones. As long as the person has not been diagnosed as infected and a TAN has not been issued to them, this is not health data. With the receipt of the TAN, the TempID in the app becomes health data that is particularly worth protecting.

b) Art. 25 GDPR provides several criteria for determining the risk posed by a processing activity, rather than by the data. These criteria include the nature, scope, circumstances, and purposes of the processing, as well as the likelihood and seriousness of the risks for the rights and freedoms of natural persons.

The type of processing refers to the form of data processing. In the present case, contact events are calculated from a large amount of distance data on the basis of heuristics, to which an exposure character is then attributed, from which serious legal and factual consequences for the persons concerned then result, such as the obligation to self quarantine.

The scope of the processing concerns any person carrying a smartphone with an installed app. Given the expectation that at least 60% of citizens will participate in this process in order for it to have an epidemiological protective effect, and the social pressure to use it, it can be assumed that this processing will affect many millions of people.

The circumstances depend on the context of the data processing. Such data processing is likely to discriminate against persons with certain characteristics, for example because they are old, or because they are infected with the Corona virus. It is also conceivable that the person responsible, or an operator of the process, or the political leadership may subsequently establish additional criteria with which further types of persons are "filtered", then controlled or managed. The purpose of the processing by the corona app is to identify and separate persons in order to be able to influence their behaviour.

The probability that persons are found using this procedure is high, and the purpose of this processing is to find certain types of persons. Furthermore, there is a high risk or probability that the procedure will filter out persons who are not corona-infected, but who nevertheless, on the basis of prima facie evidence, have to endure, at least temporarily, severe interference with their fundamental rights.

(c) As a third criterion to determining the risk level, the "Guidelines on Data Protection Impact Assessment" of the Article 29 Working Party (Article 29 Data Protection Working Party 2017) are used. In this working paper, nine different criteria are listed in order to reach a decision the risk level. Data Protection Authorities in the EU have agreed that a "likely high risk" situation exists if at least two criteria from this list are met:

1. "Evaluation or scoring, including profiling and predicting …",

2. "Automated-decision making with legal or similar significant effect …",

3. "Systematic monitoring: processing used to observe, monitor or control data subjects …",

4. "Sensitive data or data of a highly personal nature …",

5. "Data processed on a large scale …",

6. "Matching or combining datasets, for example originating from two or more data processing operations …",

7. "Data concerning vulnerable data subjects: …",

8. "Innovative use or applying new technological or organisational solutions, …",

9. "When the processing in itself prevents data subjects from exercising a right or using a service or a contract (Article 22 and recital 91) …".

(see Article 29 Data Protection Working Party 2017.)

The last seven criteria of this list apply to this procedure, as briefly outlined above.

The result of the threshold analysis is therefore that for this processing activity, a data protection impact assessment according to Art. 35 must be carried out due to a high risk for the data subjects.

# Chapter 7

# Vulnerabilities and risks

In the following we present a selection of relevant attack scenarios on the Bluetooth-based system for tracking contact events (PEPP-PT, DP-3T). The list is organized by the primary source of the attack. If an attack scenario has different potential sources, we primarily focus on the actors who are most likely to commit the attack and/or can cause the greatest damage by means of the respective attack. Within the three sections, the attack scenarios are listed in descending order of estimated risk.[1]

## 7.1 Attacks by operators, suppliers and authorities

### Attack A1: False positives – transparency and appealability of automatically imposed self-isolation

*Attacker:* Operator, authorities

The self-isolation prescribed by the corona app corresponds to an automated decision with legal consequences. Art. 22 GDPR grants users the right to challenge the decision in this case. This is particularly relevant in the case of false positive exposure measurements: for example, if person A was separated from an infected person B by the wall between their homes, the tracing system could detect a non-valid exposure event. Other realistic sources of false positives include: passengers in neighbouring train compartments; persons on either side of vehicle windows; infected residents of ground floor apartments and pedestrians on the street; forgotten smartphones, flawed bluetooth sensors and retracted false positives from laboratories.

– The inevitable danger of non-valid exposure measurements is linked to the risk that users are wrongly isolated, potentially several times in succession, with considerable economic and social consequences to those affected.

– This risk can only be controlled by an effective structure of appeal. If there is no possibility of appeal, acceptance of the system could lessen because its decisions are experienced as arbitrary. Socio-psychologically, it is conceivable that some users could deliberately seek to become infected in order to avoid the fear of isolation at any time in the future due to acquired immunity.

Due to the large number of conceivable non-valid exposure measurements through spatial proximity with a simultaneous spatial barrier between the devices (wall between apartments, window panes, etc.) as well as due to the unplanned appealability, the **risk associated with this vulnerability must be assessed as very high.**

---

[1]The list of risks makes no claim to completeness. For a complete investigation of risks, one would have to distinguish in detail between at least eight risk types. Furthermore, one would also have to address those risks that may arise additionally through the implementation of protective measures (Rost 2018b).

**Attack A2: Behavioral profiling and compliance scoring of infected users**

*Attacker:* Operator, authorities, health care system, social serice providers

The operators can use the contact histories of infected users to carry out behavioural scoring: How much contact has the person had with others? Has the person been particularly risk-taking or even negligent? To what extent has he or she complied with the applicable contact restrictions? Such behavioural scoring can be produced as part of epidemiological studies. It could however also be used for political purposes. Since the corona app is most likely to be used in a context where partial contact restrictions still apply, interest on the part of the authorities in an instrument for compliance studies is likely.

A collective and individual risk is derived from this:

– In a general variant of this attack, the scoring is anonymous: it could be used to determine statistical behavioral profiles with regard to compliance with contact requirements and risk behavior. This could easily be justified with the purpose of epidemiological investigation. However, the behavioural profiles, if correlated with other demographic and socio-economic data, may motivate selective policies in which groups of people who have been measured as being on average more willing to take risks are treated differently by future restrictions than other groups whose compliance is supposedly higher.

– The individual/specific variant of this attack results when the operators can link the tempID tokens with the identity of infected users (central variant, type 3a). In this case, authorities and politicians could pursue a number of secondary uses. On the basis of behavioural scoring, criminal prosecutions could be initiated; individual contributions to treatment costs could be levied; or decisions could be made regarding access to scarce medical resources (e.g: "whoever has been particularly negligent in his or her contact behaviour does not deserve a place on a ventilator").

The **risk of this attack is high to very high** because epidemiological evaluations are discussed and planned with seemingly wide acceptance.

**Attack A3: De-anonymisation of users based on communication data**

*Attacker:* Operators, Authorities, Organisations

In the case of the decentralized variant, the following applies: If a positively tested user loads the tempID tokens used by her app, which are initially only stored locally, onto the central server, the operators can identify the user by means of communication metadata (e.g. IP address) and device information.[2] Thus, the contact history represented by the tempID tokens can be associated with the identity of the person who tested positive. This leads to the following risks in the case of the decentralized architecture (DP-3T):

– The operators can de-anonymise positively tested persons and thus have access to health data that is personal data.

---

[2] The only protection against this would be the use of an anonymizing network like Tor, which is excluded by the DP-3T draft (DP-3T-FAQ).

- The operators can de-anonymise all contacts for persons who have tested positive, who have consequently also tested positive. With an expected 50–70% of the population ultimately catching the virus, this would represent a large part of the contact histories.

- The operators can create detailed statistical evaluations of the contact and risk behaviour of persons who test positive (including compliance with contact restriction requirements, see attack A2).

If in the context of PEPP-PT a centralized architecture, in which the tempIDs (or the seeds for their generation) are assigned from the outset by the central server, rather than the decentralized architecture (DP-3T) is implemented, then the following, much higher risks arise:

- The operators can de-anonymise all users via connection metadata.

- The operators can de-anonymise the *total* contact history of users who have tested positive.

**The decision between centralized and decentralized variants thus has considerable data protection implications.** Both variants are associated with the risk of considerable encroachments on fundamental rights, because it is in principle possible for operators to uncover the contact histories of positively tested users. In the case of the decentralised variant, however, the range of this attack is limited and de-anonymisation would require active technical and organisational deviation from the intended protocol.

**In the case of the centralized variant, there are only minor technical barriers to the de-anonymization of the users; it is even conceivable that these would be administered in a pseudonymized manner from the outset.** Changes of political will thus represent a very high risk for the protection of fundamental rights.

In the case of DP-3T, the concept is that the server does not log the connection data. Here, too, it should be specified which technical barriers stand in the way of changing this procedure.

Because the various (national) implementations within the PEPP-PT framework are given extensive freedom in the technical features relevant to this attack, the **risk of this attack must be considered internationally as very high.**

**Attack A4: Indefinite storage of data and possible later linkage with other personal data**

*Attacker: Operators, suppliers, authorities.*

It is conceivable that individual national implementations within the PEPP-PT framework will refrain from deleting the collected data after 14 days. It is also possible that corrupted or faulty implementations of DP-3T (which provides for deletion of the data) keep the data permanently (e.g. due to server backups).

If the data are not deleted after 14 days, it would be possible to link them retroactively with other data and to carry out de-anonymisation attacks.

The responsibility for preventing this attack scenario lies especially with the operators and administrators of concrete national implementations. **The risk is therefore high.**

**Attack A5: False assumptions about use cases or data processing procedures**

*Attacker:* Operators, programmers, authorities

If a user of the system tests positive, they can report this infection status at time $t$ via the system. Their contact persons will then be able to retrospectively determine their exposure risk from the time $t - n$ days, with an expected $n = 14$. In the present specifications and use cases it remains open how the infected user will proceed thereafter:

– The user could continue to use the app as before.

– The user could uninstall the app.

– The user could leave the quarantine, but not carry the smartphone with them.

It should be kept in mind that an infected user can continue to expose others; thus the history of infections, which should be recorded by the system, potentially continues. Perfect quarantine compliance is also not guaranteed. Therefore, the app would have to continue to transmit the seen tempIDs of other users on a daily basis as long as the user is considered infected. Whether this is required or intended is not technically specified.

There is a **moderate to high risk** that the integrity of the collected data will be compromised by this vulnerability, due to both the technical design and uncertainties in user behavior.

## 7.2 Attacks by private or state organizations, other interested authorities and commercial contexts

**Attack B1: Freedom restrictions when not using the app**

textitAttacker: Authorities, organisations, employers, political decision makers

Even if the use of the tracing app is voluntary, it is possible that non-use makes one subject to special restrictions on freedom of movement and contact. For example, the apps could be used as access barriers to public and private buildings, universities, schools, means of transport, administrations, police stations, etc., by requiring a corona app installed on the device to be presented upon entry.

Such a position has already been introduced into the discussion by individual actors and media reports (as of 10.04.2020): "The principle of the driving licence: those who reduce their risk potential for others with a mask and 'corona app' should not be restricted further." – This is a suggestion that the Tagesspiegel published in a guest commentary on 07.04.2020 (Schallbruch 2020).[3]

A paper by the German Federal Ministry of the Interior on the strategy for the fastest possible end of the lockdown proposes the "comprehensive contact search of positively tested persons (state 'corona detectives'/RKI initiative) [...] by means of Apps" as a central concept feature (Bundesinnenministerium 2020; *Coronakrise: Innenministerium skizziert Weg aus dem Lockdown* 2020). It describes the possiblity that the reopening of workplaces and production facilities could be made conditional on employers "having established their own (certified) protection systems". It is easy to imagine that a de facto coersion to use the tracing app could be exercised, for example, by employers solely allowing users of the app access to their buildings.

---

[3] In a similar vein, the President of the Austrian National Council, Wolfgang Sobotka, stated that a constitutional review is currently underway to determine whether the freedom of movement of persons who do not install the app can remain restricted, cf. Tremmel 2020.

It is conceivable that in such cases not only would an installed version of the app have to be presented, but also a screen showing the individual exposure risk based on the movement data of the past 14 days. In this way, it would easily be possible to check not only the installation but also the continuous use of the app.

There is a **high risk** that users could be forced to use the app even under conditions of "voluntariness". There are significant restrictions on the freedom of movement and basic rights of all those who cannot or do not want to use the app. Since not everyone owns a smartphone and the distribution of smartphones also depends on demographic and socio-economic factors, a mistreatment of non-users of the app can also lead to serious discrimination against already disadvantaged groups.

### Aggression B2: Commercial tracking

textitAttackers: Commercial operators of BT tracking infrastructure.

Shopping malls, airports, subway stations, billboards on the street, etc. are already equipped with BT-based tracking infrastructure to measure the movement of customers within a room as well as their duration of stay and return rate (e.g. at a billboard) (Kwet 2019). The MAC addresses of the BT devices (invariant and device-specific) serve as tracking tokens for these techniques.

The widespread use of a BTLE Beacon-based tracing app to combat the SARS-CoV-2 pandemic could lead to significantly more people permanently switching on the BT module of their mobile device. **It is therefore very likely** that the tracing app, if sufficiently widespread, will play into the hands of existing commercial tracking infrastructures and, as a side effect, generate a large amount of data in private hands, which can be used to study the movement behaviour of people. Not only private spaces are worth consideration here (such as retail trade), but also working environments, traffic junctions and urban areas, which are covered by privately operated electronic billboards (with BT tracking infrastructure), for example.

It is known that common MAC randomization methods from Apple can easily be circumvented and thus do not provide effective protection against this attack (Martin et al. 2019).

### Attack B3: Secondary use with central allocation of ID tokens

*Attacker:* Operators, police, authorities, political decision makers

Suppose that a user A of the system is also the subject of a criminal investigation. In principle, the contact tracing system can be used to investigate A's contact environment in order to reveal possible accomplices (see attack A3). The use of the system for such purposes would constitute a secondary use claim, which is equipped with **low technical barriers**, i.e. it is enabled or prevented primarily by the political framework. Depending on what kind of offence the person is accused of, this secondary use claim could be presented in a publicly plausible manner and easily obtained by amending the law; think in the case of a terrorist attack, for example.

The concrete risk depends on the technical implementation:

– In the case of the central architecture (here the server knows the tempIDs for all

- users), the contact history is fully discoverable if A tests positive and has sent her contact data to the server via the corona app. If A does not test positive, those of her contacts who are infected themselves and have reported this via the app are traceable.

- In the case of a decentralized architecture (DP-3T), the server only knows the tempIDs of infected users. However, by logging and evaluating communication data, the operators can de-anonymise positively tested users (see attack A3). In this case, A's contact environment is detectable if she is infected and consequently her contacts are infected. In other words, from infected suspects one can investigate all suspects who are also infected.

Similarly, secondary utilization for the persecution of refugees or, in countries not governed by the rule of law, the political persecution of persons and other minority groups is conceivable. Even if such secondary use claims do not seem politically feasible in Germany at the moment, the corona app creates an infrastructure that can be easily used for this purpose, depending on the technical design. As long as this scenario is merely prevented by political barriers, rather than *technical* ones, the **risk of this attack is considered to be very high on an international scale.** This risk is also highly relevant because some European states have recently abandoned elementary principles of the rule of law.

### Attack B4: App store and notification frameworks (Google, Apple) can be used to derive data

*Attacker:* App platforms (especially Google/Apple)

Since app platforms provide the technical and organizational infrastructure for distributing, installing, updating and uninstalling apps, as well as offer and operate exclusive notification frameworks, they can and do process the metadata generated during use. On the one hand this is done for technical reasons, on the other hand it is a central part of the business models of these companies.

This means that personal data is generated during installation, operation and uninstallation of an app on a smartphone. As the mobile applications are also based on notification frameworks such as Google's Firebase Cloud Messaging (FCM) or Apple's Push Notification Services (APN), the providers also obtain data about the communication between the app and the server. **The risk** that this data could be exploited is **high.**

On the other hand, these platforms offer a secure environment for the distribution, updating and operation of the corona app.

### Angriff B5: Platforms can trace contact events even of devices not using the corona app

*Attacker:* Smartphone Platforms (e.g. Google Android)

In 2018 it became known that mobile phones with Google Android receive Bluetooth beacons even if the Bluetooth module is switched off. If the user has activated the "Google Location History" function, Android systems regularly transmit the registered beacons of other devices to a Google server, provided that the "High accuracy location mode" or "Bluetooth scanning" is switched on at the same time (Yanofsky 2018). This is a configuration that is found among a large number of users.

It is possible that this functionality also applies to the BTLE beacons of the corona app. This means that all Android phones configured as described above, even those which do not have the corona app installed, might register the tempIDs of nearby corona app users – even those using Apple devices – and send them to a Google server.[4] This results in considerable data protection risks:

- Google could thus comprehensively record the contact history of eData protection authorityData protection authorityven those users who do not use the Corona App themselves. The only requirement is that the user has activated the "Google Location History" function and at the same time switched on the "High accuracy location mode" or "Bluetooth Scanning".

- If the TempIDs of a user who tested positive are published via the CA system, Google could use these records to determine which of the received TempIDs belong to infected users and which Android devices had contact with them. Google might thus be able to determine exposure events across all devices, even for users who have not installed the corona app themselves.

- By combining this information with other data, such as the Google account of the Android user, Google could in most cases also associate this information with names, phone numbers, and other identifying information. Thus, a deanonymization of infected users could be done by Google.

- By triangulating the registered BTLE beacons of several Android devices in close proximity, it can be assumed that Google could also map tempIDs to the sending devices, even before a user publishes his/her TempIDs after testing positive.

Because the technical infrastructure for this attack already exists and in principle scales to the entirety of all Android users, the risk associated with this attack is very high. In particular, it should be pointed out that this vulnerability also affects uninvolved parties, that is, users who do not install the corona app themselves.

## 7.3 Attacks by hackers, trolls, stalkers and individuals

### Attack C1: Large-scale bluetooth hacking

*Attacker:* Third parties, hackers, police, intelligence services

A BT-based corona app forces users to permanently switch on the BT module of their mobile device (see attack B2). The BT implementations of both Android and iOS are known to have numerous vulnerabilities,[5] some of which even allow the unnoticed execution of arbitrary code on the attacked device. Large-scale, quasi-forced use of BT on the terminals of a majority of the population could encourage attackers to exploit these vulnerabilities. One conceivable scenario would be automated break-ins into the devices of any users at a location where the attacker and the target are in close proximity for a short time, such as in residential buildings, public transport or at waiting points.

In order to be effective, the corona app must be used by a large part of the population. To achieve this level of saturation, the app will not be able to insist on the latest security updates or use the latest operating system versions. A corona app that is effective in terms of its distribution would have to have relatively low requirements regarding the software environment. As a result, the use of the corona app will expose devices with known weaknesses in BT implementation. The **risk of this attack is considered moderate** because the attacks are locally limited and therefore scale poorly.

---

[4]See also: https://github.com/DP-3T/documents/issues/222 (visited:2020-04-23).
[5]See the references listed in the "Security" section in Privacy International 2020.

### Attack C2: Injection of false infection events

*Attacker:* Third parties, hackers, police, intelligence agencies

Generally, neither the app nor the server can check whether the positive testing of a user, which is the trigger for the transmission of the tempID tokens to the server, has actually taken place. It would be possible that a user passes the TAN issued for data transmission to *another* user or that users obtain TANs illegally.

It is therefore conceivable that individual actors may inject false infection events into the system with the consequence that other users may be incorrectly classified as exposed. There are different classes of attack, which also depend on the measures chosen to ensure the integrity of the processing activity:

– If the server cannot and/or does not check the tempID tokens, then any tempIDs in the value range can be sent to the server. As a result, any app users will potentially be assigned a risk score beyond the threshold and will be mistakenly treated as exposed.

– If the server only accepts authentic tempID tokens, then there must be a procedural element in the processing activity that checks the link between the data upload and the positive test, for example, by involving a healthcare institution. On this basis, an authentication token such as an anonymous Credential would have to be assigned, which can be checked after the tempID tokens have been uploaded. In this case, the attack can still be carried out in densely populated regions by a relatively small group of attackers moving around for several days, visiting places where a large number of contact events occur.

As far as the authentication tokens for the data upload (TANs) are assigned in a highly decentralized way, e.g. by individual medical institutions, we expect organized attackers to obtain such authentication tokens relatively easily, in order to upload large amounts of tempID tokens to the server allegedly based on positive test results. As far as the authentication tokens are centrally issued, the problem is only postponed: attackers do not have to try to get authentication tokens directly, but could try to obtain false medical certificates about their infection status, which can then serve as credentials to obtain authentication tokens. Because a certain amount of criminal energy and organization is needed for this attack, and because the attack scales only moderately, its **risk is moderate.**

## Chapter 8

# Determination of protection measures for processing activities

Based on the risk sources (e.g. attackers), threats (e.g. attacks or malfunctions) and legal requirements (see Chapter 5) described in the previous chapter, further requirements are set out below in order to address the high protection needs of personal data that is logically passed on to the components of the processing activities.

The protection measures are each marked with an "R" to make a reference to the risks they address. For example, the risk posed by attack A4 is marked with [R:A4].

The protection measures are specified for the superordinate processing procedure as a whole, systematically including the individual processing activities.

All requirements are formulated as MUST, SHOULD or MAY phrases. The word MUST indicates that requirements are necessary measures without which processing must not take place. The word SHOULD indicates that requirements are necessary measures for which there may, in certain circumstances, be valid reasons for not complying with them, but the full extent of the non-compliance must be understood, carefully weighed and the decision and reasons must be clearly and comprehensively documented. The word MAY denote requirements that are completely optional.

With regard to the present processing activities, at least the following modules of the Standard Data Protection Model (SDM-V2) shall be implemented for the processing procedure and its components (see Section 4):

| target object | module |
| --- | --- |
| CA, CA Server | module 11 "Data Retention" |
| processing procedure, processing activities, processes, data, formats, specialist applications, IT services, operational processes, communication relationships | module 41 "Planning and Specification" |
| processing procedure, processing activities, processes, data, formats, specialist applications, IT services, operational processes, communication relationships | module 42 "Documentation" |
| processes, specialist applications, IT services, operational processes, communication relationships | module 42 "Logging" |
| processing procedure, processing activities, processes, data, formats, specialist applications, IT services, operational processes, communication relationships | module 50 "Separation" |
| processes, specialist applications, IT services, processes | module 60 "Deleting and Destroying" |
| organisation | module 80 "Data Protection Management" |

## 8.1 Overarching protection measures for the entire processing procedure

**M 0.1 Data Protection Management (transparency, intervenability) [R:A1 R:A4].**
Data protection management (DPM) MUST be in place to identify, monitor and control data protection requirements (see legal requirement on page 44). Implementation notes are included in SDM-V2.

A DPM MUST be set up as a data protection management system (DPMS) when there is a high risk to the rights and freedoms of a person, which in turn is subject to continuous improvement according to the state of the art. Mature management systems show *Key performance indicators (KPI)*. In our context a desirable KPI to measure the maturity of the DPMS regarding the processing procedure could be to show the frequency of checks made for the case that Health-TempIDs are uploaded without logging.

**M 0.2 Use of a decentralized architecture (unlinkability) [R:A3 R:B3]**
To minimize the risks of deanonymization or the creation of social graphs, a decentralised architecture MUST be preferred.

**M 0.3 Specification of the processing procedure (transparency, integrity) [R:A5]**
The processing activities MUST be completely planned and specified (see SDM module "Planning and Specification"). All relevant use cases MUST be recorded and described.

For high-risk processing procedures, the planning phase of their architecture must be carefully designed. This includes using a proven project management method such as the waterfall method or PRINCE2. Project management methods SHOULD in turn be framed by a general quality management assurance framework. Standardised processes must be established at the interface between the organisational structure and IT systems, which can deal with problems and change requirements with reference to the organisational structure. Here, ITIL in particular is an established standard in the European Union (Rost and Welke 2020).

A test concept MUST be in place to document test phases that are subject to controlled release as well as protocols that reflect design decisions.

**M 0.4 Protection against (non-)use-related discrimination (unlinkability) [R:B1]**
The publication of the CA MUST be accompanied by legal and factual measures to ensure that users do not have to disclose the status of the CA or the existence of the CA on their own devices to others. Exceptions MAY be made, for example, for medical staff to enforce home quarantine, even for employers, by means of sick leave. The aim of these regulations is to ensure the purpose limitation of the CA. CA-based access barriers to public and private buildings, universities, schools, transport, administration, police stations, etc. MUST be eliminated. It MUST be examined to what extent the necessary measures are to be located outside their control, for example, in the legislative area.

Purpose limitation is essentially enforced by separating databases, IT systems and services as well as sub-processes. In this context, the conditioning of the personal reference by means of pseudonymisation of the TempID, the anonymisation of the Health-TempIDs uploaded to the CA server, and their conversion to "non-identifying infection-indicating data (NIID)" is essential. It MUST be examined how to enforce the operation of different IT systems on the CA server side and their organisational separation.

The verifiability of a processing activity depends on the preparations from the planning phase with regard to the documentation of the properties of the components involved, with which nominal value and actual values can be determined, as well as the logging, with which system events in the past can be traced. If the risk is high, the logging MUST have a certified time stamp and precautions MUST be taken to ensure that the identifiers for the instances involved are unique and meaningful to allow human readability. Logging MUST then be set up so that log data is not stored on the production machines, but on a dedicated log server. Any checking activities of log data MUST also be logged. These serve the controller to prove the effectiveness of the protection measures.

### M 0.5 Secure app development (integrity) [R:B4]

Care SHOULD be taken to use a secure development environment for app development.

Here, it is important to ensure that what was said above in the context of planning a processing activity. Agile development methods can also be used for the development of software for high-risk processing activities, but special attention should be paid to legal requirements. The top priority is not the rapid development of functionalities, but compliance of data protection and law.

### M 0.6 Secure distribution of the app (integrity) [R:B4]

Checking mechanisms SHOULD be provided to allow the user to verify the integrity of the app after downloading.

### M 0.7 Secure use of the app (integrity, confidentiality, unlinkability) [R:B4]

An IT usage policy SHOULD be provided for the users. In addition, information SHOULD be provided on supported devices and operating system versions that have been tested by the controller and checked for possible data flows.

The app SHOULD be part of the IT security audit according to BSI IT-Grundschutz, especially to secure the communication with the CA server through encryption and authentication. It MUST be evaluated to what extent an audit should also cover all relevant operating system components and even devices on which the app is not installed. It MUST also be evaluated how effective this protective measures actually is. The cooperation of the platform operators (Google, Apple) is essential for this.

### M 0.8 Balancing between the degree of dissemination and secure use of CA (integrity, availability, confidentiality) [R:C1]

On the market and even more in use are various smartphone operating systems and operating system versions, many of them not up to date and therefore with known security holes, including an insecure Bluetooth stack. When using the CA, a permanent activation of the Bluetooth module is required. The operator MUST check and weigh up to what extent a high degree of distribution of the CA is actually much more important than guaranteed secure operation in every use. In the first case, the CA MUST support as many smartphone operating system versions as possible; in the second case, the possibility of installing the CA must be limited to current operating system versions that are still supported.

**M 0.9 Provision of a data protection policy (transparency)**
A data protection statement in accordance with Art. 21 para. 1 GDPR (see also page 60) MUST be offered to the user before use. It MUST be available at least in the national language, MAY be prepared in all the languages frequently spoken in the area of use, but SHOULD also be available in the languages of particularly vulnerable groups if this differentiation appears to be useful.

The data protection policy SHOULD contain precise information about what the exact parameters of the contact and risk value calculation are.

**M 0.10 Protection measures for specialist applications, CA and CA server and TAN management (confidentiality, integrity)**
Due to the high protection needs, the security measures MUST be certified according to BSI IT-Grundschutz or ISO 27001.

**M 0.11 Protection measures for all IT services (confidentiality, integrity)**
The IT services used include smartphone, BTLE service, database server, server operating system, hardware, data center and the Internet connection. Due to the high protection needs, the security measures for these IT services MUST be certified according to BSI IT-Grundschutz or ISO 27001.

**M 0.12 Protection measures for all communication relationships via TCP/UDP (confidentiality, integrity)**
The transport link between server and client MUST be secured end-to-end according to the state of the art. If necessary, a public key infrastructure with a medium or high validation level must be used for this.

**M 0.13 Establishing the auditability of the source code (transparency)**
The source code for the CA and the CA server MUST be publicly available. It MUST always be available at least to the supervisory authorities in an up-to-date form.

**M 0.14 Audit-proof logging of software changes (transparency, integrity)**
The change of software versions must be recorded in an audit-proof and publicly viewable manner.

**M 0.15 Evaluation of the use of app platforms (confidentiality, unlinkability) [R:B4]**
The controller MUST evaluate to what extent the data protection risks outweigh the IT security advantages or vice versa when using app platforms. In any case, legal, technical and/or organizational measures MUST be taken to prevent the usage data of the platforms from being linked with the data of the CA.

**M 0.16 Evaluation of the use of Bluetooth in the context of related functions used by smartphone operating system platforms (confidentiality, unlinkability) [R:B5]**
The controller MUST investigate whether common smartphone operating systems (Android, iOS) and their Bluetooth modules transfer CA data to the platform operators (see R:B5). If this is the case, then the extent of the risky context of processing MUST be estimated and reassessed together with the supervisory authorities. If the risk is too high, for example because more than 10,000 smartphones would be affected, processing MUST be suspend. In the context of R:B5, it can be assumed that the configuration of the Android system that exposes the devices is the default setting of the operating system and is therefore used by a large number (possibly the majority) of Android users. If at the same time it cannot be excluded that the system in this configuration forwards the BTLE beacons used by the corona app to a Google server, the risk would undeniably be too high.

The protection measures that can be assigned to individual processing activities (PA) are described below.

## 8.2 Protection measures for the processing activity "App-side processing of contact events"

### M A.1 Separate storing in the app (unlinkability) [R:A2]
The storing of TempIDs sent on the one hand and received on the other hand on the smartphone MUST be done separately to prevent the reconstruction of contact events.

Important events, such as the time the app was launched, SHOULD be logged.

### M A.2 Store contact events only locally (unlinkability) [R:A2 R:A4]
If contact events are stored so that own and foreign TempIDs can be linked, this record MUST always remain on the device.

### M A.3 Unlinkability of TempIDs (Unlinkability) [R:A2] [R:A4]
It MUST be impossible to associate different TempIDs of the same user. For example, they MUST not be based on a mathematical seed that allows for later interlinking.

The source code of the generating function (usually the CA) MUST be accessible and audited (see M 0.13).

### M A.4 Protection against third-party tracking (unlinkability, confidentiality) [R:B2]
BTLE beacons MUST be generated by an appropriate method so that the format and data do not allow conclusions to be drawn about the users and the equipment used. It MUST be ensured that the transmitted TempIDs are changed at appropriate time intervals.

### M A.5 Backup of personal TempIDs (availability)
The controller SHOULD check to what extent backups of the self-generated, personal TempIDs are possible according to their protection requirements.

## 8.3 Protective measures for the processing activity "Authorization of upload, anonymisation, temporary storage and dissemination of positive infection status"

### M B.1 Protection of anonymity (confidentiality, integrity, unlinkability) [R:A3 R:B3]
It MUST be ensured that no protocols exist on the server side that allow the de-anonymization of users.

There MUST be strong access control for the server.

It MUST be ensured that access control is implemented on the server using state-of-the-art disc encryption.

Access control to the server MUST be implemented on the basis of a two-man rule to enforce legal and organizational separation. This separation MUST be implemented with technical support. Access MUST also be implemented via multi-factor authentication.

The use of an appropriate anonymisation service on the client side MUST be examined. The lack of an anonymisation service MUST be compensated by an over-organisational separation, e.g. by having an extra-organisational institution as part of the access control. Otherwise the start of this processing activity MUST be suspended.

The transport link between server and client MUST be secured end-to-end and authenticated acording to the state-of-the-art.

It MUST be examined to what extent the measures to be taken with regard to the protection of anonymity can provide adequate protection or whether a new legal form for the organisational separation of server access must be created by the legislator (see also page 55).

### M B.2 Planning and specification of the procedure after notification of infection (transparency, integrity) [R:A5]

It MUST be planned and specified what happens after an infection notification is sent to the server by the user. It MUST be specified what function the CA performs after the message. The user MUST be given meaningful instructions in clear and simple language on how to proceed after the message and when the process is complete for them. It SHOULD be examined whether it is appropriate to continue reporting the infection status via TempIDs in order to keep the contact history up to date.

### M B.3 Publish server statistics (transparency) [R:A4]

To ensure data protection-friendly server operation, the server MAY regularly publish aggregated status data, such as the total number of non-identifying infection-indicating data (NIID).

### M B.4 Personal use of the TAN (integrity) [R:C2]

It MUST be ensured that the TAN can only be used by the corresponding person. One possibility is to issue TAN lists with two halves of TANs to doctors, who then give the first half of the TAN to the patients during medical treatment, the second half of the TAN only after the laboratory has reported a positive diagnosis and the doctors inform the patients of the diagnosis. During the telephone call, the patients then have time to use the entire TAN to initiate the transmission of the Health-TempIDs to the server. Since the first half of the TAN is exchanged among those present, third parties cannot reconstruct the entire TAN. Since the entire TAN is only valid during the telephone call, it can most likely only be used by the infected person. Since the person's doctor does not look at the smartphone, he or she cannot determine whether the TAN has been used or not.

A different procedure for ensuring exclusively personal use MAY be developed and implemented. It MUST be proven that this other procedure has at least equally good data protection properties.

**M B.5 Protection against spoofing/tampering/impersonation of TempIDs (integrity, confidentiality, unlinkability) [R:C2]**
It MUST be ensured that the TempIDs are unique worldwide and that the value range cannot be efficiently enumerated (Enumeration Attack). The use of an anonymous verification system (see e.g. Camenisch and Lysyanskaya 2001) SHOULD be eximined.

**M B.6 Planning and specification of the TAN management process (transparency, integrity, confidentiality, unlinkability, availability) [R:C2]**
The process for creating, transmitting, using and releasing TANs MUST be planned and specified. It MUST be ensured that the interlinking of TANs and Health-TempIDs remains strictly confidential.

**M B.7 Retrieval, withdrawal, or deletion of TempIDs already sent to the server (intervenability, transparency) [R:A1]**
Since laboratory tests can also be faulty (false negatives, false positives, swapped samples, subsequent correction of the measurement, more precise tests with negative results) or if the user decides not to release his or her data anymore, it MUST be technically possible to delete already uploaded TempIDs from the server. This MUST happen within a legal regulation for deadlines. In addition the users MUST be informed about the consequences and the further procedure.

**M B.8 Delay of upload in case of an unavailable CA server (availability)**
In the event that the CA server or an Internet connection is not available during the upload process, this process SHOULD be delayed until the server is available again. Termination MUST be avoided.

**M B.9 Inclusion of backups for deletion periods (unlinkability)**
Since the server will create backups for availability reasons, the controller MUST examine to what extent deletion requests from a user must also affect these secondary storages.

**M B.10 Measures to ensure high availability of the server (availability)**
The controller MUST examine to what extent high availability requirements are to be placed on the server and how these are to be implemented if necessary.

## 8.4 Protective measures for the processing activity "Decentralized contact tracing"

**M C.1 Planning and specification of the process for influencing decisions by CA (transparency, intervenability, integrity)**
A process MUST be planned and specified that allows the user to influence the CA's decision.

It MUST be possible for the user to undo the decision. Users MUST be informed of the legal remedies available and be able to access the information about them at any time.

There MUST be a single point of contact (SPoC). The SPoC MUST be retrievable in the CA.

**M C.2 Backup of the received TempIDs (availability)**
The controller SHOULD check to what extent backups of received TempIDs are possible according to their protection needs.

## Chapter 9

# Recommendations for data controllers on how to design the processing and implement the protections measures identified

In order to fulfill the requirements of the GDPR, it is recommended that the following protective measures be taken by the controller of the smartphone app to detect risky contacts with COVID-19 infected persons:

1. An appropriate legal basis must be established and in this respect responsibilities have to be defined. Framing the processing as "voluntarily" and conducting it on the basis of consent does not meet the requirements of data protection law, particularly since doubts exist concerning the voluntary and conscious status of the user. Instead, legal bases are to be created to meet these requirements, in particular in terms of purpose limitation, anonymisation, the deletion concept, and the data protection management. Here it is vital to not only pay attention to the technical specifications of the app alone, but to the entire processing procedure including its interfaces, for example the integration into the planned electronic reporting procedure. Undesirable technical and social side-effects influencing the exercise of fundamental rights must be taken into account as well. Those side-effects also indirectly affect the acceptance of the processing procedure. Thus, it must be ensured that third parties do not have access to the app and its output on the smartphones of the affected persons. The legislature must adopt a regulation in accordance with § 14 (8) IfSG (German infection protection act) to specify technical requirements in conformity with data protection.

2. With respect to an identified or identifiable natural person, two situations in the entire process chain are particularly sensitive; namely the creation and storage of the TempIDs and uploading the Health TempIDs of CV infected persons to the server and storing them as NIIDs. These neuralgic operations must be designed as follows:

    a. When creating the TempIDs in the app, it must be ensured that there is **no connection between TempIDs** and it must never be possible to make one. A concrete TempID must therefore neither be derived from the previous one nor from a common seed. The TempIDs must be stored in the app in a way that makes it impossible to determine retrospectively in which order they were created and saved.

    b. The server(s)' operator must employ an **effective separation method**, which transforms the health TempIDs from the apps of COVID-19 infected

persons into non-identifying infection-indicating data (NIID) on the server. This transformation must be legally, organizationally and technically ensured (Podlech 1976). **Legally** the operator must be an independent body which must not have an interest in the data and is protected from obligations to disclose data,including security authorities. **Organizationally** the contoller must strategically establish a mix structure together with the operator ensuring functional differentiation or the informational separation of powers to be enforced within the organization. For example, jurisdiction and court management will work together and yet separately in the court organization. The operator must have a data protection management which allows for the separation to be effectively enforced and maintained in a verifiable manner. **Technically** the operator has to set up the separation process so that uploads of the health TempIDs are logged neither on the server nor in the operators' network. Furthermore, the upload of health TempIDs between apps and servers must be end-to-end encrypted and protected by upstream anonymization proxies (e.g. Tor). For data protection control the separation method must be continuously verified by the responsible data protection authority.

The IT Security of all IT components used in the entire process chain, including the interaction with doctors and public health departments must be certified according to BSI IT-Grundschutz or according to ISO 27001. Especially aspects of ensuring availability, in particular of server (infrastructures), the authentication of the IT components involved and the confidentiality of communication connections must be considered according to their high protection needs.

3. Accompanying the publication of the app it *must* be ensured in law and in fact that users have to disclose neither the status of the app nor the mere existence on a device to third parties. An exception could be made for medical personnel to enforce home quarantine by using sick leave towards employers. The aim of these regulations is to ensure the purpose limitation of the app. Access controls for public and private buildings, universities, schools, means of transport, public administrations, police stations etc. based on disclosing the app status must be prevented.

4. Before the app gets published, a comprehensive investigation of the software and the overall system *must* be conducted and published by an independent body. Particular attention must be paid to the risks that arise from interactions with operating system components, which are potentially relevant also for non-users of the app (see attack scenario B5 in chapter 7). This point could fail if large platform companies (Google, Apple) are unwilling to cooperatre, or if legal or factual obstacles prevent securing data protection requirements (see for example the U.S. CLOUD Act or the PRISM program of the U.S. National Security Agency). If, for example, it cannot be ruled out that platforms may read contact events and perform matching with infected persons, the *termination* of the corona app projects would be the appropriate consequence.

# Abbreviations

| | |
|---|---|
| **Art.** | Article |
| **BDSG** | Federal Data Protection Act |
| **BMG** | Federal Ministry of Health |
| **BT** | Bluetooth |
| **BTLE** | Bluetooth Low Energy |
| **CA** | CA type 3 dezentral / Corona App |
| **COVID-19** | Corona Virus Disease 2019 |
| **CV** | Corona Virus |
| **DP-3T** | *Decentralized Privacy-Preserving Proximity Tracing* (Projekt) |
| **DPM** | Data Protection Management |
| **DPMS** | Data Protection Management System |
| **DPO** | Data Protection Officer |
| **DPIA** | Data Protection Impact Assessment |
| **DSK** | Conference of the Independent Federal and State Data Protection Authorities in Germany |
| **FIfF** | Forum InformatikerInnen für Frieden und gesellschaftliche Verantwortung |
| **GDPR** | General Data Protection Regulation |
| **GPS** | Global Positioning System |
| **NIID** | Non-identifying Infection-Indicating Data |
| **IfSG** | Protection against Infection Law (Germay) |
| **KPI** | Key-Performance-Indicator |
| **PA** | Processing Activity |
| **PEPP-PT** | *Pan-European Privacy-Preserving Proximity Tracing* (Framework) |
| **RKI** | Robert Koch-Institut |
| **SDM** | Standard Data Protection Modell |
| **SPoC** | Single-Point-of-Contact |

*Abbreviations*

**TCP**        Transmission Control Protocol

**TAN**        Transactionnumber

**TempID**      Temporary Pseudonymous Identifier

**UDP**        User Datagram Protocol

# Glossary

## Types of data

A. Locally on the sending smartphone with the CA
   a. TempID (character string, calculated and stored by the smartphone)
   b. TempID token (character string, sent as Bluetooth user data, can contain hardware address data)
   c. Health TempID (like TempID plus TAN authorization, sent to server)

B. CA Server
   a. Health TempID (like TempID plus TAN authorization, received from client)
   b. The separation process must take place here.
   c. Infection-indicating date without information relating to an identified or identifiable natural person (sent to apps as updates), abbreviation: "NIID"

C. Locally on the receiving smartphone with CA
   a. Infection-indicating date without information relating to an identified or identifiable natural person (received from server)
   b. Contact date (measured foreign TempID, time duration, signal strength profile over time)
   c. Match = exposure event (calculated)
   d. Risk score
   e. Infection warning (binarization of the risk score)

For the fundamental relationship between data, their context and the relationship to the concept of information, a core element of the separation process, see also Dreyfus 1972, S. 197 ff.

## Data protection terms

**Data protection** Data protection has the function of placing the relationships between organisations and individuals under normative and operational conditions when organisations dominate these relationships. The balance of power asymmetry, which is required by fundamental rights, is effectively enforced, with the help of data protection law, by the design of processing activities and the operation of protective measures on the side of the organisations. Data protection must be distinguished from *Safety* (ensuring the operation of technical components), *IT security* (ensuring business processes through protective measures on the part of the organizations in the interest of the organizations) and *Privacy* (protection of persons with measures either on the part of the persons themselves or for personal data on the part of the organizations in the sense of IT security).

**processes** A processing procedure can be based on several organizational, technical or personnel processes that are necessary for the operation of the procedure.

Other designations for this are support process, operating process, subprocess.

**Controller** According to Art. 4 GDPR (definitions), this means the natural or legal person, authority, institution or other body which alone or jointly with others determines the purposes and means of processing personal data.

**Processing procedure, processing activity**

> "For the purposes of this Regulation [...] 'processing' means any operation or set of operations which is performed on personal data or on sets of personal data, whether or not by automated means, such as collection, recording, organisation, structuring, storage, adaptation or alteration, retrieval, consultation, use, disclosure by transmission, dissemination or otherwise making available, alignment or combination, restriction, erasure or destruction." (Art. 4 para. 2 GDPR)

The individual processing activity of personal data is the object of the DSFA and therefore its starting point. The processing activity also takes place in context. Although the processing context is part of the analysis and the impact assessment, the technical and organisational measures derived from it, which encompass the context, are generally applied solely to the object.

**Processing procedure** The processing procedure describes the execution of a processing or processing activity and thus has a process character. While the processing procedure essentially answers the question of how a processing is to be carried out, the processing is essentially determined by its functional purpose, which it gives to the processing procedure.

According to the Standard Data Protection Model, the concept of processing procedure is defined as follows: "The term 'processing procedure' is used to describe complete data processing operations. Data processing includes in particular any collection, storage, modification, transmission, blocking, erasure, use, anonymisation, pseudonymisation and encryption of personal data. A processing procedure describes a formalised, repeatable sequence of these above-mentioned data processing steps for the implementation of a technical task or a business process. It does not matter whether they are carried out manually or with the aid of information technology. A processing procedure is always characterized by its intended purpose and is thus distinguished from other procedures."

The legal definition is in Art. 4 para. 2 DSGVO.

In organisational life, procedures are typically known as "specialist processing procedures" or "business processes".

**Processing procedure analysis** Breakdown of the processing procedure by processes and related aspects (1) data and data formats, (2) systems and interfaces, (3) processes and roles

**Process** A transaction is the sum of data, systems and processes.

A process is a processing step within a processing procedure. An operation can be linked to other operations to form an operation sequence. A series of operations is complete and can be understood as a processing procedure if the purpose of the processing procedure can be realized.

**Purpose** Functional justification for a processing or processing activity within a social context.

## State of health/disease

**Infected** Person is infected with the virus (may still be latent, not contagious; may already be contagious without symptoms; may be COVID-19 infected) Infection ends when the disease is cured (symptoms subside, possibly immunisation) or with death.

**Sick (of CoViD-19)** Person is infected and has symptoms.

**Contagious** Acutely infected person in the course of the disease, in which the infection is also contagious.

**Positive tested** Infected person has been tested positive for the virus using suitable methods (e.g. PCR test for viral RNA) and is thus acutely infected.

**Exposed, potentially infected** Person has been exposed to an infectious agent in an epidemiologically relevant way (for example, sufficiently long, sufficiently close).

*Glossary*

# Changelog

**Version 1.6 – 29. April 2020**
- Inhaltlich: Aufgrund eines technischen Fehlers ausgelassene Satzteile wieder eingefügt.
- Formal: PDF-Metadaten, FIfF logo added to front page

**Version 1.5 – 24. April 2020**
- Inhaltlich: Angriff B5 hinzugefügt.
- Formal: Rechtschreibkorrekturen, übersehene Kommentare entfernt und Aktualisierung der Danksagung

**Version 1.4 – 18. April 2020**
- Inhaltlich: eine Referenz eingefügt (CDU2019)
- Formal: Rechtschreibkorrekturen, Verbesserung der Formulierungen, übersehene Kommentare entfernt.

**Version 1.3 – 17. April 2020**
- Inhaltlich: keine
- Formal: Rechtschreibkorrekturen, Lesbarkeitsverbesserungen

**Version 1.2 – 16. April 2020**
- Inhaltlich: keine
- Formal: Rechtschreibkorrekturen, übersehene Kommentare entfernt und Aktualisierung der Danksagung

**Version 1.1 – 14. April 2020**
- Inhaltlich: keine
- Formal: Rechtschreibkorrekturen und aktualisierung Danksagung

**Version 1.0 – 14. April 2020**
- initiale Veröffentlichung

*References*

# Index

*INDEX*